\documentclass[10pt,aps,twocolumn,prc,superscriptaddress,showpacs,nofootinbib,noshowkeys,floatfix,preprintnumbers]{revtex4}
\usepackage[dvips]{graphics,graphicx}
\usepackage{amsmath, amssymb}
\usepackage{multirow}
\usepackage{longtable}
\usepackage{color}
\usepackage[normalem]{ulem}  
\newcommand{\LL}{$\Lambda\Lambda$}
\newcommand{\DBLL}{\Delta B_{\Lambda\Lambda}}
\newcommand{\reff}{r_\mathrm{eff}}
\newcommand{\fm}{\mathrm{fm}}
\newcommand{\MeV}{\mathrm{MeV}}
\newcommand{\GeV}{\mathrm{GeV}}

\renewcommand\sout{\bgroup \color{red} \ULdepth=-.5ex \ULset}
\begin{document}
\title{Lambda-Lambda interaction from relativistic heavy-ion collisions}
\date{\today}
\preprint{YITP-14-67}
\author{Kenji Morita}
\email{kmorita@yukawa.kyoto-u.ac.jp}
\affiliation{Yukawa Institute for Theoretical Physics, Kyoto University,
Kyoto 606-8502, Japan}
\affiliation{Frankfurt Institute for Advanced
Studies, Ruth-Moufang-Str. 1, D-60438 Frankfurt am Main, Germany}
\affiliation{Institute of Theoretical Physics, University of Wroclaw,
PL-50204 Wroc\l aw, Poland}
\author{Takenori Furumoto}
\affiliation{National Institute of Technology, Ichinoseki College, Ichinoseki, Iwate 021-8511, Japan}
\author{Akira Ohnishi}
\email{ohnishi@yukawa.kyoto-u.ac.jp}
\affiliation{Yukawa Institute for Theoretical Physics, Kyoto University,
Kyoto 606-8502, Japan}
\begin{abstract}
 We investigate the two-particle intensity correlation function of
 $\Lambda$ in relativistic heavy-ion collisions. We find that the
 behavior of the $\Lambda\Lambda$ correlation function
 at small relative momenta is fairly sensitive to the interaction potential
 and collective flows.
%
 By comparing the results of different source functions and potentials,
 we explore the effect of intrinsic collective motions
 on the correlation function.
We find that the recent STAR data gives a strong constraint
 on the scattering length and effective range of $\Lambda\Lambda$ interaction
 as,
 $-1.8~\mathrm{fm}^{-1} < 1/a_0 < -0.8~\mathrm{fm}^{-1}$ 
 and 
 $3.5~\mathrm{fm} < r_\mathrm{eff} < 7~\mathrm{fm}$, respectively,
if $\Lambda$ samples do not include feed-down contribution from
 long-lived particles.
 We find that feed-down correction for $\Sigma^0$
 decay reduces the sensitivity  of the correlation function to the detail of the $\Lambda\Lambda$
 interaction. As a result, we obtain a weaker constraint $1/a_0 < -0.8$ fm$^{-1}$.
 Implication for the signal of existence of $H$-dibaryon is discussed.
 Comparison with the scattering parameters obtained from 
 the double $\Lambda$ hypernucleus may reveal in-medium effects
 in the $\Lambda\Lambda$ interaction.
\end{abstract}
\maketitle

\section{Introduction}

Hyperon-hyperon interaction plays an important role in various aspects
of modern nuclear physics such as 
hypernuclear, 
exotic particle, 
neutron star,
and strange matter
physics.
Current most precise information on \LL\ interaction is 
obtained from the double $\Lambda$ hypernuclear mass~\cite{Nagara,FG,Hiyama},
and it is closely related to the existence of the $S=-2$ dibaryon 
(the $H$ particle).
In neutron star core,
hyperons have been believed to emerge and to soften the equation of
state~\cite{HyperEOS}.
While recent observation of massive neutron stars disfavors
admixture of strange hadrons~\cite{Demorest},
hypernuclear physics data suggest hyperon admixture
at $\rho_{\scriptscriptstyle B}=(2-4)\rho_0$~\cite{HyperEOS}.
Deeper understanding of hyperon-hyperon interaction 
may help solving this massive neutron star puzzle.
At very high densities,
hyperon superfluid could be continuously connected 
to the color superconductor,
where the three flavors ($uds$) and colors ($rgb$) are entangled.

The existence of the $H$ particle has been one of the long-standing problems
in hadron physics.
In 1977, Jaffe pointed out that double strange dibaryon 
made of 6 quarks ($uuddss$) may be deeply bound 
below the $\Lambda\Lambda$ threshold 
due to the strong attraction from color magnetic interaction
based on the bag model calculation~\cite{Jaffe}.
Deeply bound $H$ was ruled out by the existence of double $\Lambda$ hypernuclei.
A double $\Lambda$ hypernucleus $^{~~6}_{\Lambda\Lambda}\mathrm{He}$ 
was found to decay weakly (Nagara event),
and the observed energy of $^{~~6}_{\Lambda\Lambda}\mathrm{He}$
is 
$7.25~\MeV(=B_{\Lambda\Lambda})$
below the $^4\mathrm{He}+\Lambda\Lambda$ threshold~\cite{Nagara}.
Since $^{~~6}_{\Lambda\Lambda}\mathrm{He}$ should decay to
$^4\mathrm{He}+H$ if the mass of $H$ is $M_H < 2M_\Lambda-B_{\Lambda\Lambda}$,
the deeply bound $H$ was ruled out.

The $H$ particle is again attracting much attention
due to recent theoretical and experimental efforts.
%
Recent lattice QCD calculations have demonstrated that
$H$ appears as the bound state around the flavor SU(3) limit,
and they also suggest the possibility for $H$ to appear
as a bound state or resonance pole~\cite{LQCD}.
Experimentally,
KEK-E224 and KEK-E522 experiments~\cite{E224,E522} demonstrated that 
\LL\ invariant mass spectrum is enhanced in the low energy region
between $\Lambda\Lambda$ and $\Xi N$ thresholds,
compared with the phase space estimate
and the classical transport model calculations~\cite{Nara}.
Enhancement of the invariant mass spectrum just above the threshold
implies 
the final state interaction effects of \LL\ attraction.
KEK-E522 data also show the bump structure around 10 MeV above
the \LL\ threshold.  This bump cannot be explained solely
by the final state interaction effects, but it is not significant enough 
($\sim 2\sigma$) to claim the existence of a resonance pole~\cite{E522}.

It is evident that we need higher statistcs data
to obtain more precise information on \LL\ interaction,
and eventually to conclude the existence/non-existence of the $H$ pole.
%
Higher statistics data will be available from the future J-PARC experiments
on double $\Lambda$ hypernuclear observation
and \LL\ invariant mass measurement,
as proposed in J-PARC E42 experiment.
%
Recently, an alternative possibility to access the information on
hadron-hadron interactions has been explored 
in heavy-ion collisions 
at Relativistic Heavy Ion Collider (RHIC)
and Large Hadron Collider (LHC).
The large hadron multiplicity, which is achived by hadronization from
the quark-gluon plasma (QGP), makes it possible
to look at correlations
between hadrons with good experimental statistics. In particular, 
the intensity correlation of identical particles in relative momentum
space is known as 
the Hanbury-Brown and Twiss (HBT) or Goldhaber-Goldhaber-Lee-Pais (GGLP) effect 
to give information on the size of the emission source
through the (anti-)symmetrization of the two-boson (fermion) wave
function.

The HBT effect of stable hadrons, particularly pions, has been used to
estimate the source sizes created in relativistic nucleus-nucleus
collisions. 
On the one hand, effects of interaction between two of those emitted
particles on extracting source size can be
absorbed into the chaoticity parameter,
when the system has a large size compared with the interaction range.
On the other hand, one expects substantial effects of the interaction on the correlation 
function of identical particles of which interaction is sufficiently
strong in the range comparable to the effective source size
 ~\cite{Lednicky,Bauer:1992}.
This implies that one may be able to use the correlation function
to obtain information on the interaction between two identical particles,
even if those particles are unstable.
For \LL\ pair, this idea is not new.
It was proposed in '80s that we can fix resonance parameters,
when the source size is small~\cite{GreinerMuller1989}.
The correlation at low relative momenta was proposed to be useful
to discriminate the sign of the scattering length $a_0$,
provided that the source size is large~\cite{OHNSA};
When \LL\ has a bound state ($a_0>0$),
the scattering wave function must have a node
at $r \simeq a_0$
in order to be orthogonal to the bound state wave function,
then we may find the suppression of the correlation.
The vertex detectors at RHIC 
have enabled us to really obtain the \LL\ correlation data 
in heavy-ion collisions with a good signal-to-noise ratio
by choosing weakly decaying $\Lambda$ off the reaction 
point~\cite{STAR,Shah2012}.

In this paper, along with the above expectation, 
we investigate the interaction between two $\Lambda$ baryons.
Several hyperon-nucleon ($YN$) and hyperon-hyperon ($YY$) interaction models
have been proposed so far by constraining parameters 
from limited number of $YN$ scattering experiments,
flavor symmetries, and hypernuclear data.
We calculate the \LL\ correlation functions with
those interaction potentials. Given a model source function relevant
for Au+Au collisions at $\sqrt{s_{NN}}=200$~GeV, we discuss the
modification of the correlation function due to the interaction and
collective flow effects, then show that how the behavior the correlation
function constraints the nature of the interaction.
A preliminary report
of the present work can be found in Ref.~\cite{Hyp2012}. In this paper,
we present a detailed systematic analysis with the updated experimental
data of $\Lambda\Lambda$ correlation.

In Ref.~\cite{STAR}, the STAR Collaboration has reported and
analyzed their experimental data of $\Lambda\Lambda$ correlation 
with the Lednick\'{y} and Lyuboshitz analytical model~\cite{Lednicky}
which incorporates the
effect of the $\Lambda\Lambda$ interaction in terms of the effective
range and the scattering length
together with
the intercept (chaoticity) parameter $\lambda$ 
and normalization as fitting parameters.
Moreover, it has been shown that the inclusion of a residual correlation 
as an additional gaussian term responsible to the high-momentum tail
gives a better description of the data.
In this paper, we focus on effects of the $\Lambda\Lambda$ interaction
through the modification of the wave function 
and the deformation of the emission source function owing to the
collective flow which takes place in relativistic heavy-ion collisions.
We will show that the $\Lambda\Lambda$ correlation data measured by the STAR Collaboration
including the intercept and the residual correlation can be explained 
with some of the recent $\Lambda\Lambda$ potentials
and flow parameters constrained by the single particle spectrum of
$\Lambda$, if we assume feed-down correction is negligible.
We will then examine effects of $\Sigma^0$ feed down correction on the
obtained constraints and discuss their interplay with a residual correlation. 

This paper is organized as follows.
In Sec. \ref{Sec:LLint},
we briefly summarize models of the \LL\
interaction. In Sec.~\ref{sec:source}, we introduce the two-particle
correlation function with the final state interaction and show the
general property based on a simple source model. Intrinsic effects in 
relativistic heavy-ion collisions are discussed in Sec.~\ref{sec:flow}.
We discuss feed-down correction and residual correlations in Sec.~\ref{sec:feeddown}.
We also discuss possible implications for the $H$ particle 
in Sec.~\ref{sec:discussion}. Section \ref{sec:conclusion} is devoted to
concluding remarks.

\section{$\Lambda\Lambda$ interaction potential}
\label{Sec:LLint}

\begin{table*}
\caption{$\Lambda\Lambda$ potentials.
The scattering length ($a_0$) and effective range ($r_\mathrm{eff}$)
are fitted using a two-range gaussian potential,
$V_{\Lambda\Lambda}(r)=V_1 \exp(-r^2/\mu_1^2)+V_2 \exp(-r^2/\mu_2^2)$.
}\label{tbl:potential}
\begin{tabular}{l|rr|rrrr|l}
\hline
\hline
Model 
& $a_0$ (fm) & $r_\mathrm{eff}$ (fm)
& $\mu_1$ (fm) & $V_1$ (MeV)
& $\mu_2$ (fm) & $V_2$ (MeV)
& Ref.
\\
\hline
ND46      &$  4.621$& 1.300&1.0 &$-144.89$  &0.45& 127.87&\protect{\cite{ND}}    $r_c=0.46$ fm\\
ND48      &$ 14.394$& 1.633&1.0 &$-150.83$  &0.45& 355.09&\protect{\cite{ND}}    $r_c=0.48$ fm\\
ND50      &$-10.629$& 2.042&1.0 &$-151.54$  &0.45& 587.21&\protect{\cite{ND}}    $r_c=0.50$ fm\\
ND52      &$ -3.483$& 2.592&1.0 &$-150.29$  &0.45& 840.55&\protect{\cite{ND}}    $r_c=0.52$ fm\\
ND54      &$ -1.893$& 3.389&1.0 &$-147.65$  &0.45&1114.72&\protect{\cite{ND}}    $r_c=0.54$ fm\\
ND56      &$ -1.179$& 4.656&1.0 &$-144.26$  &0.45&1413.75&\protect{\cite{ND}}    $r_c=0.56$ fm\\
ND58      &$ -0.764$& 6.863&1.0 &$-137.74$  &0.45&1666.78&\protect{\cite{ND}}    $r_c=0.58$ fm\\
NF42      &$ 3.659$ & 0.975&0.6 &$ -878.97$ &0.45&1048.58&\protect{\cite{NF}}    $r_c=0.42$ fm\\
NF44      &$23.956$ & 1.258&0.6 &$-1066.98$ &0.45&1646.65&\protect{\cite{NF}}    $r_c=0.44$ fm\\
NF46      &$-3.960$ & 1.721&0.6 &$-1327.26$ &0.45&2561.56&\protect{\cite{NF}}    $r_c=0.46$ fm\\
NF48      &$-1.511$ & 2.549&0.6 &$-1647.40$ &0.45&3888.96&\protect{\cite{NF}}    $r_c=0.48$ fm\\
NF50      &$-0.772$ & 4.271&0.6 &$-2007.35$ &0.45&5678.97&\protect{\cite{NF}}    $r_c=0.50$ fm\\
NF52      &$-0.406$ & 8.828&0.6 &$-2276.73$ &0.45&7415.56&\protect{\cite{NF}}    $r_c=0.52$ fm\\
NSC89-1020&$-0.250$ & 7.200&1.0 &$  -22.89$ &0.45&  67.45&\protect{\cite{NSC89}} $m_\mathrm{cut}=1020$ MeV\\
NSC89-920 &$-2.100$ & 1.900&0.6 &$-1080.35$ &0.45&2039.54&\protect{\cite{NSC89}} $m_\mathrm{cut}=920$ MeV \\
NSC89-820 &$-1.110$ & 3.200&0.6 &$-1904.41$ &0.45&4996.93&\protect{\cite{NSC89}} $m_\mathrm{cut}=820$ MeV \\
NSC97a    &$-0.329$ &12.370&1.0 &$ -69.45$  &0.45& 653.86&\protect{\cite{NSC97}} \\
NSC97b    &$-0.397$ &10.360&1.0 &$ -78.42$  &0.45& 741.76&\protect{\cite{NSC97}} \\
NSC97c    &$-0.476$ & 9.130&1.0 &$ -91.80$  &0.45& 914.67&\protect{\cite{NSC97}} \\
NSC97d    &$-0.401$ & 1.150&0.4 &$-445.77$  &0.30& 373.64&\protect{\cite{NSC97}} \\
NSC97e    &$-0.501$ & 9.840&1.0 &$-110.45$  &0.45&1309.55&\protect{\cite{NSC97}} \\
NSC97f    &$-0.350$ &16.330&1.0 &$-106.53$  &0.45&1469.33&\protect{\cite{NSC97}} \\
Ehime 	  & $-4.21$ & 2.41 &1.0 &$-146.6$   &0.45& 720.9 &\protect{\cite{Ehime}} \\
fss2  	  & $-0.81$ & 3.99 &0.92&$-103.9$   &0.41& 658.2
			 &\protect{\cite{fss2}}  \\
ESC08     &$-0.97 $ &3.86  &0.80& $-293.66$ &0.45& 1429.27 &\protect{\cite{ESC08}}\\
\hline
\hline
\end{tabular}
\end{table*}

We examine several models of \LL\ interaction
proposed so far by using \LL\ correlation in heavy-ion collisions.
%
Since experimental information on \LL\ interaction is limited,
\LL\ correlation data are useful to constrain \LL\ interaction.
\LL\ interaction is known to be weakly attractive from the \LL\ bond energy
in $^{~~6}_{\Lambda\Lambda}\mathrm{He}$,
$\DBLL=
B_{\Lambda\Lambda}(^{~~6}_{\Lambda\Lambda}\mathrm{He})
-2 B_{\Lambda}(^{5}_{\Lambda}\mathrm{He}) \simeq 1.01~\mathrm{MeV}$~\cite{Nagara}.
From $\DBLL(^{~~6}_{\Lambda\Lambda}\mathrm{He})$,
the scattering length and the effective range 
in the \LL\ $^1\mathrm{S}_0$ channel 
are suggested as
$(a_0, \reff)=(-0.77~\fm, 6.59~\fm)$~\cite{FG}
or 
$(a_0, \reff)=(-0.575~\fm, 6.45~\fm)$~\cite{Hiyama},
but in principle one cannot determine two low energy scattering parameters
from a single observed number of $\DBLL$.
For example, while $\reff$ values are very similar, $a_0$ are different
by 25-30 \% in the above two estimates.
Low energy \LL\ scattering parameters are useful
to distinguish the models of baryon-baryon ($BB$) interaction.
The long-range part of the $NN$ interaction is dominated
by the one pion exchange potential, which roughly determines
the low energy behavior of $NN$ scattering.
By comparison,
$\Lambda$ particle is isoscalar and there is no one-pion exchange in
\LL\ interaction. Thus the low energy \LL\ scattering parameters, 
such as the scattering length $a_0$ and the effective range
$r_\mathrm{eff}$, are more sensitive to the $BB$ interaction models.

There are several types of \LL\ interactions proposed so far.
Meson exchange model \LL\ interactions~\cite{ND,NF,NSC89,NSC97,ESC08,Ehime}
have a long history of studies.
Nijmegen group have provided several versions of $NN$, $YN$ and $YY$
interactions, 
model D (ND)~\cite{ND},
model F (NF)~\cite{NF},
soft core (NSC89 and NSC97)~\cite{NSC89,NSC97},
and extended soft core (ESC08)~\cite{ESC08}.
These interactions have been widely used
in hypernuclear structure calculations~\cite{FG,Hiyama}.
Compared with $NN$ and $YN$ interactions,
we have larger uncertainties in $YY$ interaction
and there are some rooms to vary model parameters.
We regard the hard core radius $r_c$ in hard core models (ND and NF)
and the cutoff mass $m_\mathrm{cut}$ in NSC89
as free parameters.
In the case of NSC97, there are several versions (NSC97a-f)
having different spin dependence in $\Lambda N$ interaction,
and \LL\ scattering parameters
would help discriminating these versions.

Ehime potential is a boson exchange \LL\ potential~\cite{Ehime},
whose strength is fitted to the old double $\Lambda$
hypernuclear bond energy,
$\DBLL=4~\MeV$~\cite{Old-DoubleL}.
 Since this value is proven to be too large,
Ehime potential is now known to be too attractive.
Even though, it would be valuable to examine the \LL\ correlation
with more attractive potential than usually considered.

 \begin{figure}
  \includegraphics[width=3.375in]{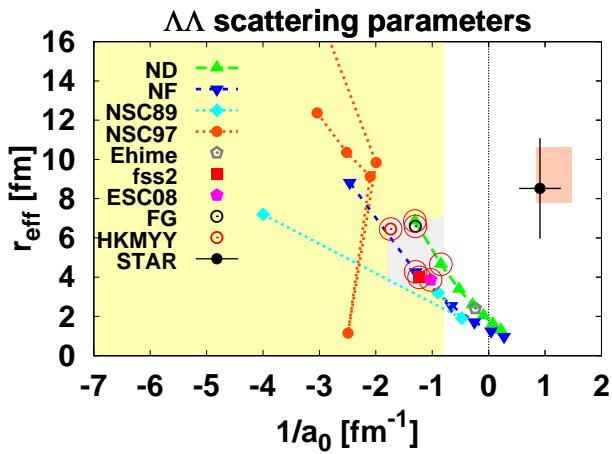}
  \caption{$\Lambda\Lambda$ interactions
  and scattering parameters in the $(1/a_0, \reff)$ plane.
  The \LL\ interations favored by the \LL\ correlation data 
  without feed-down correction are marked with big circles.
  The thin big and thick small shaded area correspond to the favored regions of 
  scattering parameters with and without feed-down correction, respectively,
  which show stable and small $\chi^2$ minimum. (See text.)
  The results of the analysis by the STAR collaboration
  is shown by the filled circle~\protect\cite{STAR}, together with
  systematic error represented by the surrounding shaded region. 
  }
  \label{fig:AR-sel}
 \end{figure}

The quark model $BB$ interactions have a merit
that the Pauli principle between quarks and the one-gluon exchange
give rise to the short range repulsion,
which seems to be consistent with other $NN$ interaction models.
At the same time,
in order to describe the medium- and long-range part of the $BB$ interaction,
we need to take account of the meson exchange between quarks or baryons.
There are several quark model $BB$ interactions 
which include the meson exchange effects.
We adopt here the fss2 model~\cite{fss2},
as a typical quark model interaction.
This interaction is constructed for the octet-octet $BB$ interaction
and describes the $NN$ scattering data at a comparable precision 
to meson exchange potential models.
For fss2, we use a phase-shift equivalent local potential
in the two range gaussian form~\cite{fss2},
derived by using the inversion method
based on supersymmetric quantum mechanics~\cite{SB}.


Low energy scattering parameters of the \LL\ interactions considered here
are summarized in Table \ref{tbl:potential}. In Fig.~\ref{fig:AR-sel},
we show the scattering parameters ($1/a_0$ and $r_\mathrm{eff}$) of the
$\Lambda\Lambda$ interactions under consideration.
These scattering parameters characterize the low-energy scattering phase shift
in the so-called shape independent form as
\begin{align}
k\cot\delta = -\frac{1}{a_0} + \frac12 r_\mathrm{eff} k^2 + \mathcal{O}(k^4)
\ .
\end{align}
For negatively large $1/a_0$, the attraction is weak and the phase shift
rises slowly at low energy. When we go from left to right in the figure,
the interaction becomes more attractive and a bound state appears when
$a_0$ becomes positive.
%
We have parameterized the boson exchange \LL\ interactions,
described above in two-range gaussian potentials
\begin{align}
V_{\Lambda\Lambda}(r)=
 V_1 \exp(-r^2/\mu_1^2)
+V_2 \exp(-r^2/\mu_2^2)
\ ,\label{eq:TRG}
\end{align}
then fit the low energy scattering parameters, $a_0$ and $r_\mathrm{eff}$.

\begin{table*}[!t]
\caption{$\Lambda\Lambda$ potentials from Nagara event.
The scattering length ($a_0$) and effective range ($r_\mathrm{eff}$)
are fitted using a three-range gaussian potential,
$V_{\Lambda\Lambda}(r)=V_1 \exp(-r^2/\mu_1^2)+V_2 \exp(-r^2/\mu_2^2) + V_3 \exp(-r^2/\mu_3^2)$.
}\label{tbl:potential_nagara}
\begin{tabular}{l|rr|rrrrrr|l}
\hline
\hline
Model 
& $a_0$ (fm) & $r_\mathrm{eff}$ (fm)
& $\mu_1$ (fm) & $V_1$ (MeV)
& $\mu_2$ (fm) & $V_2$ (MeV)
& $\mu_3$ (fm) & $V_3$ (MeV)
& Ref.
\\
\hline
HKMYY &$-0.575$ &6.45  &1.342& $-10.96$ &0.777& $-141.75$ &0.35&2136.6 &\protect{\cite{Hiyama}}\\
FG     &$-0.77 $ &6.59  &1.342& $-21.49$ &0.777& $-250.13$ &0.35&9324.0 & \protect{\cite{FG}}\\
\hline
\hline
\end{tabular}
\end{table*}

In addition to the $\Lambda\Lambda$ potentials listed in Table
\ref{tbl:potential}, we also examine the potentials used in
Refs.~\cite{FG} (by Filikhin and Gal; FG) and \cite{Hiyama} 
(by Hiyama, Kamimura, Motoba, Yamada and Yamamoto; HKMYY)
with the three-range Gaussian fit
given in those references. The parameters are summarized in Table
\ref{tbl:potential_nagara}. 

Before closing the section, we note that the coupling effects with 
$\Xi N$ and $\Sigma\Sigma$ channels are effectively incorporated in the
present treatment, since the coupling modifies the low energy scattering
parameters of $\Lambda\Lambda$ and we use the low energy phase shift
equivalent potential. Also in Refs.\cite{FG,Hiyama}, the coupling
effects with $\Xi N$ is included in the $\Lambda\Lambda$ potential. The
explicit coupling effect on the $\Lambda\Lambda$ correlation would be an
interesting subject, but is out of the scope of this paper.

\begin{widetext}

\section{$\Lambda\Lambda$ correlation function with interaction effects}
\label{sec:source}

  \subsection{Formalism}

  A relevant formulation of the two-proton correlation function
  is given in \cite{Gong:1991} which solidates the formula in
  \cite{Koonin:1977}. Here we apply the formula to $\Lambda\Lambda$ correlation.
   Then, for a given relative wave function $\Psi_{12}$, the correlation
  function defined as two-particle distribution
  $W_2(\boldsymbol{k_1},\boldsymbol{k_2})$ normalized by the
  one-particle distributions $W_1(\boldsymbol{k_i}) (i=1,2)$ can be
  expressed in terms of one-particle phase space density $S(x,\boldsymbol{k})$ as
  
  \begin{align}
   C_2({\boldsymbol Q},{\boldsymbol K}) &= \frac{W_2(\boldsymbol{k_1},\boldsymbol{k_2})}{W_1(\boldsymbol{k_1})W_1(\boldsymbol{k_2})}\\
   &=\frac{\displaystyle \int d^4
    x_1 d^4 x_2 S(x_1,\boldsymbol{K})S(x_2,\boldsymbol{K})
    |\Psi_{12}(\boldsymbol{Q},\boldsymbol{x_1}-\boldsymbol{x_2}-(t_2-t_1)\boldsymbol{K}/m)|^2}{
    \displaystyle \int d^4x_1 d^4 x_2
    S(x_1,\boldsymbol{k_1})S(x_1,\boldsymbol{k_2})}\label{eq:c2}
  \end{align}
  where $\boldsymbol{K}=(\boldsymbol{k_1}+\boldsymbol{k_2})/2$ and
  $\boldsymbol{Q}=\boldsymbol{k_1}-\boldsymbol{k_2}$ are the average and the
  relative momentum of the two identical particles, respectively. Since
  the expression for the two-particle distribution is derived for small
  $\boldsymbol{Q}$, one can also put 
  $\boldsymbol{k_1}\simeq  \boldsymbol{k_2} \simeq \boldsymbol{K}$ in
  $S(x,\boldsymbol{k_i})$ in the denominator. 
\end{widetext}

As $S(x,\boldsymbol{k})$ represents the one-particle phase space
distribution of $\Lambda$, effects of interaction are embedded in the
relative wave function $\Psi_{12}(\boldsymbol{Q},\boldsymbol{r})$.
As we are considering the effect of \LL\ interaction 
through the potential $V(r)$, the relative wave function
given by solving the Schr\"{o}dinger equation is time independent.
The factor $-(t_2-t_1)\boldsymbol{K}/m$, with $m$ being the mass of
$\Lambda$, is added to the relative coordinate
in $\Psi_{12}$ to take into account the different emission time of two
$\Lambda$ particles. 

The two-particle wave function respects the anti-symmetrization for
the two identical fermions. For the non-interacting 
spin-singlet (spin-triplet) case,
the spatial part of the wave function is symmetric (anti-symmetric)
with respect to the exchange of the two particle position,
\begin{align}
 \Psi_{s} &= \frac{1}{\sqrt{2}}
  (
    e^{i\boldsymbol{k_1\cdot x_1}+i\boldsymbol{k_2\cdot x_2}}
  + e^{i\boldsymbol{k_2 \cdot x_1 } +i\boldsymbol{k_1\cdot x_2}}
  ) \nonumber\\
 & = \frac{1}{\sqrt{2}}e^{2i \boldsymbol{K\cdot X}}
  (e^{i\boldsymbol{Q\cdot r}/2} + e^{-i \boldsymbol{Q \cdot r}/2})
\\
 \Psi_t &= \frac{1}{\sqrt{2}}
  (
   e^{i\boldsymbol{k_1\cdot x_1}+i\boldsymbol{k_2\cdot x_2}}
  -e^{i\boldsymbol{k_2 \cdot x_1 } +i\boldsymbol{k_1\cdot x_2}}
  ) \nonumber\\
 & = \frac{1}{\sqrt{2}}e^{2i \boldsymbol{K\cdot X}}(e^{i
 \boldsymbol{Q\cdot r}/2}-e^{-i \boldsymbol{Q \cdot r}/2}) 
\end{align}
where we have introduced the center-of-mass coordinate
$\boldsymbol{X}=(\boldsymbol{x_1}+\boldsymbol{x_2})/2$ and relative one
$\boldsymbol{r}=\boldsymbol{x_1}-\boldsymbol{x_2}$.

With the interaction described by a potential $V(r)$,
we assume here that only the $s$-wave is modified.
The two-particle wave function in the spin singlet state
is represented as, with the solution of the Schr\"{o}dinger equation in
the $s$-wave $\chi_Q(r)$,
\begin{align}
 \Psi_s &= \sqrt{2}
\left[\cos(\boldsymbol{Q\cdot r}/2)+\chi_Q(r)-j_0(Qr/2)
\right]\ ,
\end{align}
where $j_0(Qr/2)$ is the spherical Bessel function at zeroth order and 
we omit $\boldsymbol{X}$ dependent part as they give unity in
$|\Psi_{s,t}|^2$. 
The spin-averaged total wave function squared is now given by
\begin{align}
 |\Psi_{12}|^2&= \frac{1}{4}|\Psi_s|^2 + \frac{3}{4}|\Psi_t|^2\\
 &= 1-\frac{1}{2}\cos(\boldsymbol{Q\cdot r}) +
 \left[ \chi_Q(r)-j_0\left(\frac{Qr}{2} \right) \right] \nonumber \\
 & \times\cos\left(\frac{\boldsymbol{Q\cdot r}}{2}\right) +\frac{1}{2}\left[
 \chi_Q(r)-j_0(Qr/2) \right]^2\label{eq:wf}
\end{align}

If we neglect the interaction, $\chi_Q(r)=j_0(Qr/2)$, 
only the first and second terms remain
to give the free HBT correlation which has an
intercept $1/2$ at $\boldsymbol{Q}=0$. This reflects 
symmetrization of the spatial wave function in the spin-singlet channel
and the anti-symmetrization of the spatial wave function due to the
Pauli principle in the spin-triplet channel, 
$2 \times 1/4 + 0 \times 3/4 = 1/2$. 
On the other hand, it has been known that the bosonic correlation function gives the
intercept value two. Thus, we expect $C(\boldsymbol{Q}=0) < 1/2$ for
repulsive interactions and $C(\boldsymbol{Q}=0) > 1/2$ for attractive
ones.
Comparisons of the full correlation function with the free correlation
function  give direct information on the effect of the interaction, as
we shall see below.

\subsection{Static spherically symmetric source}

 \begin{figure*}[ht]
  \includegraphics[width=2.2in]{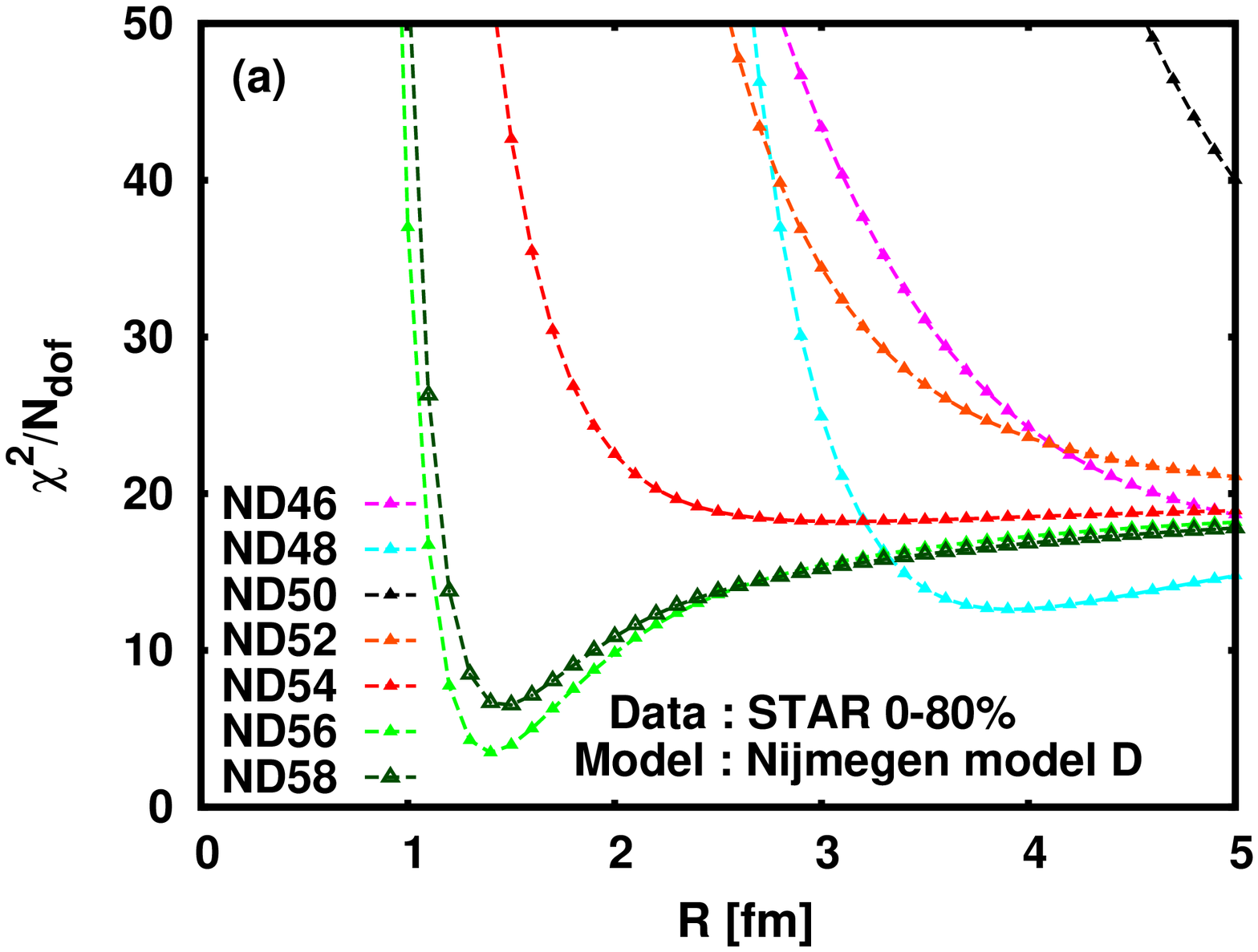}%
  \includegraphics[width=2.2in]{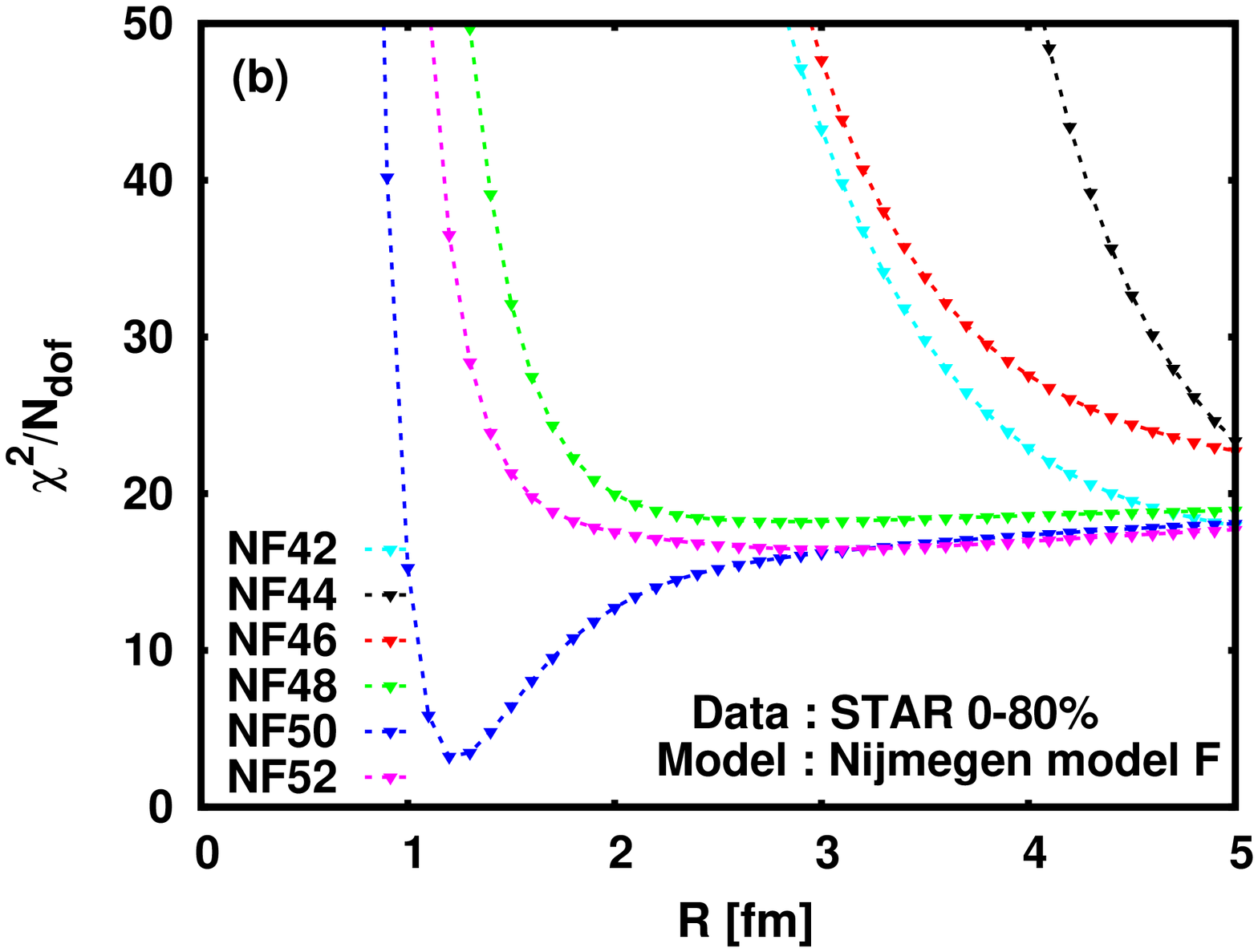}%
  \includegraphics[width=2.2in]{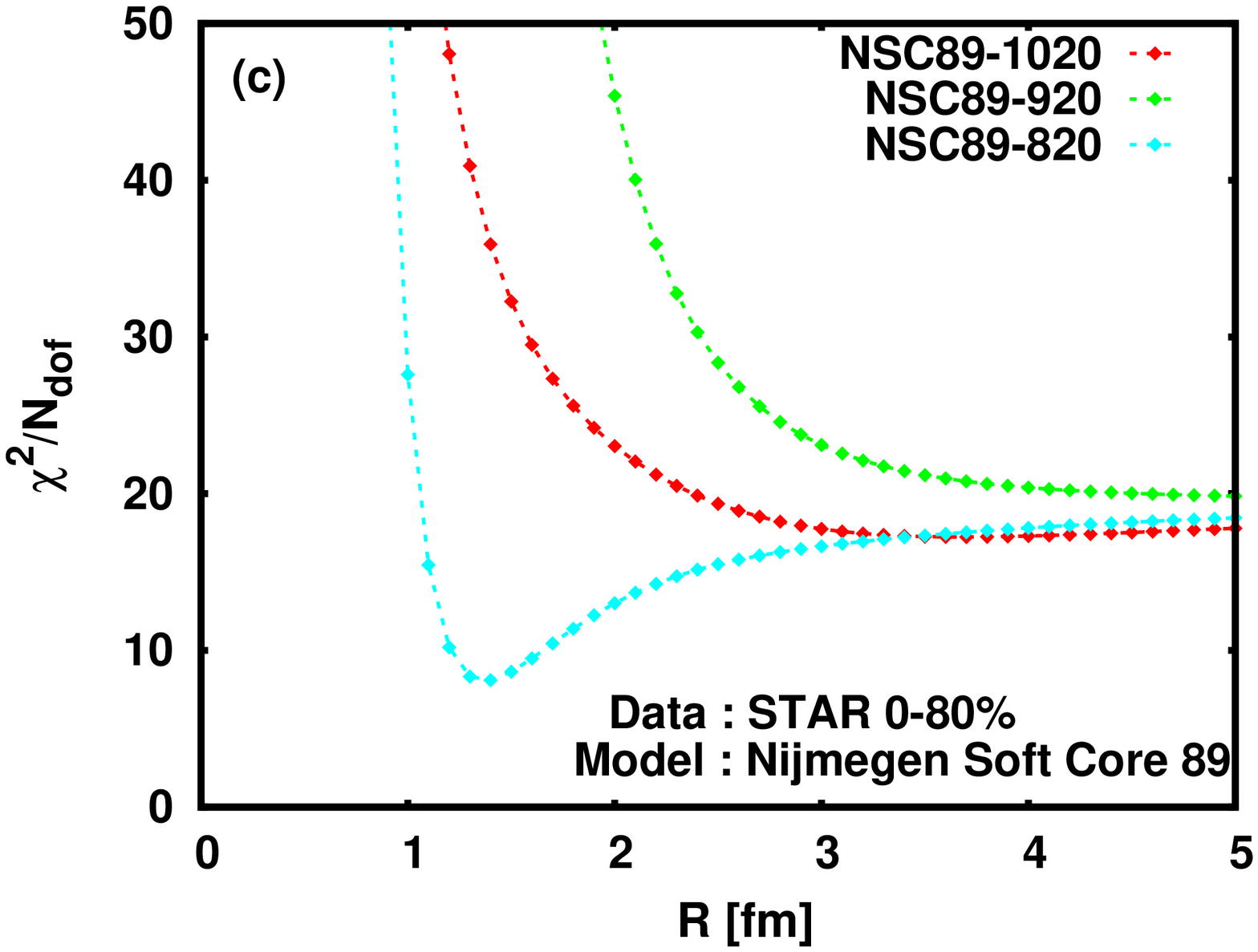}%
\\
  \includegraphics[width=2.2in]{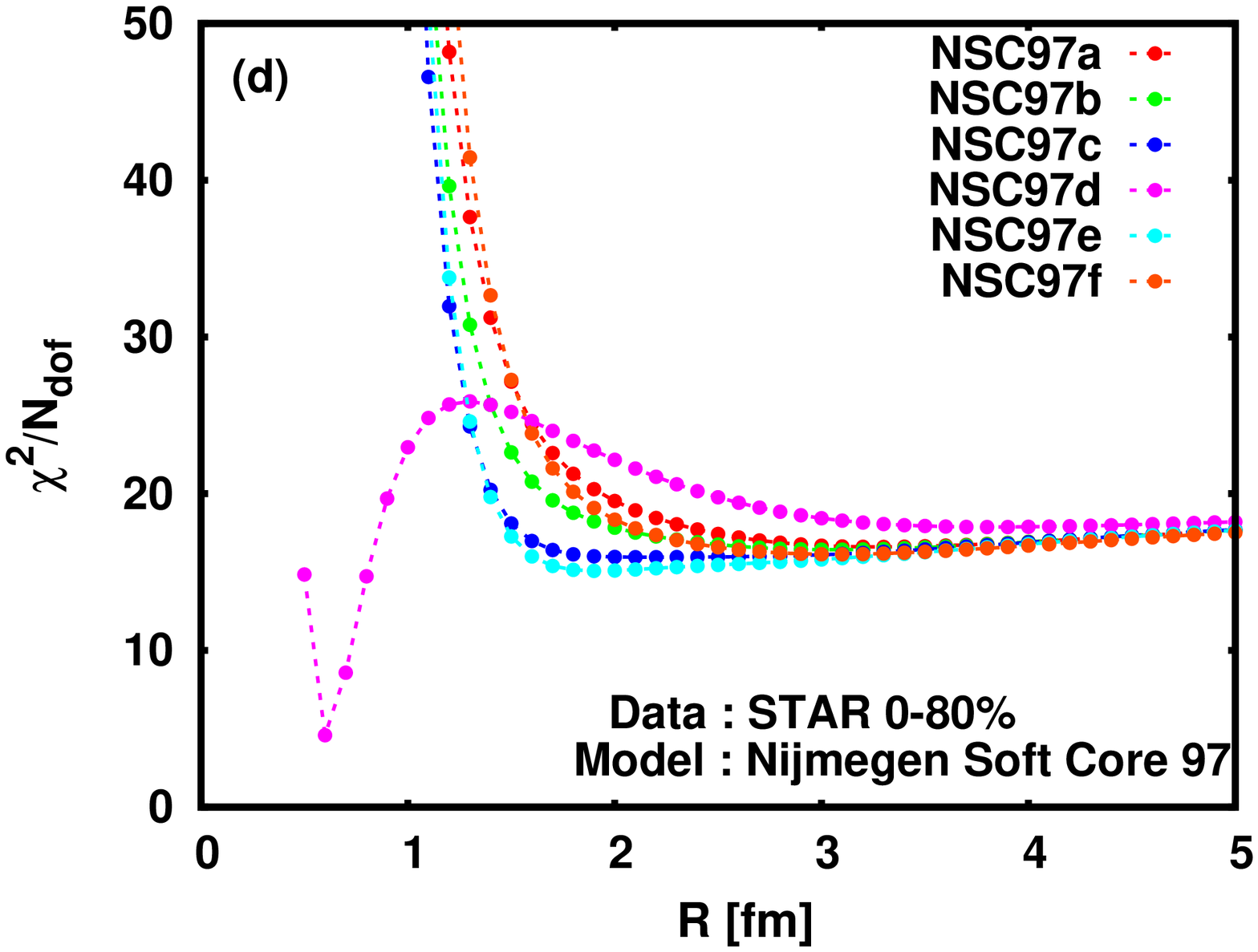}%
  \includegraphics[width=2.2in]{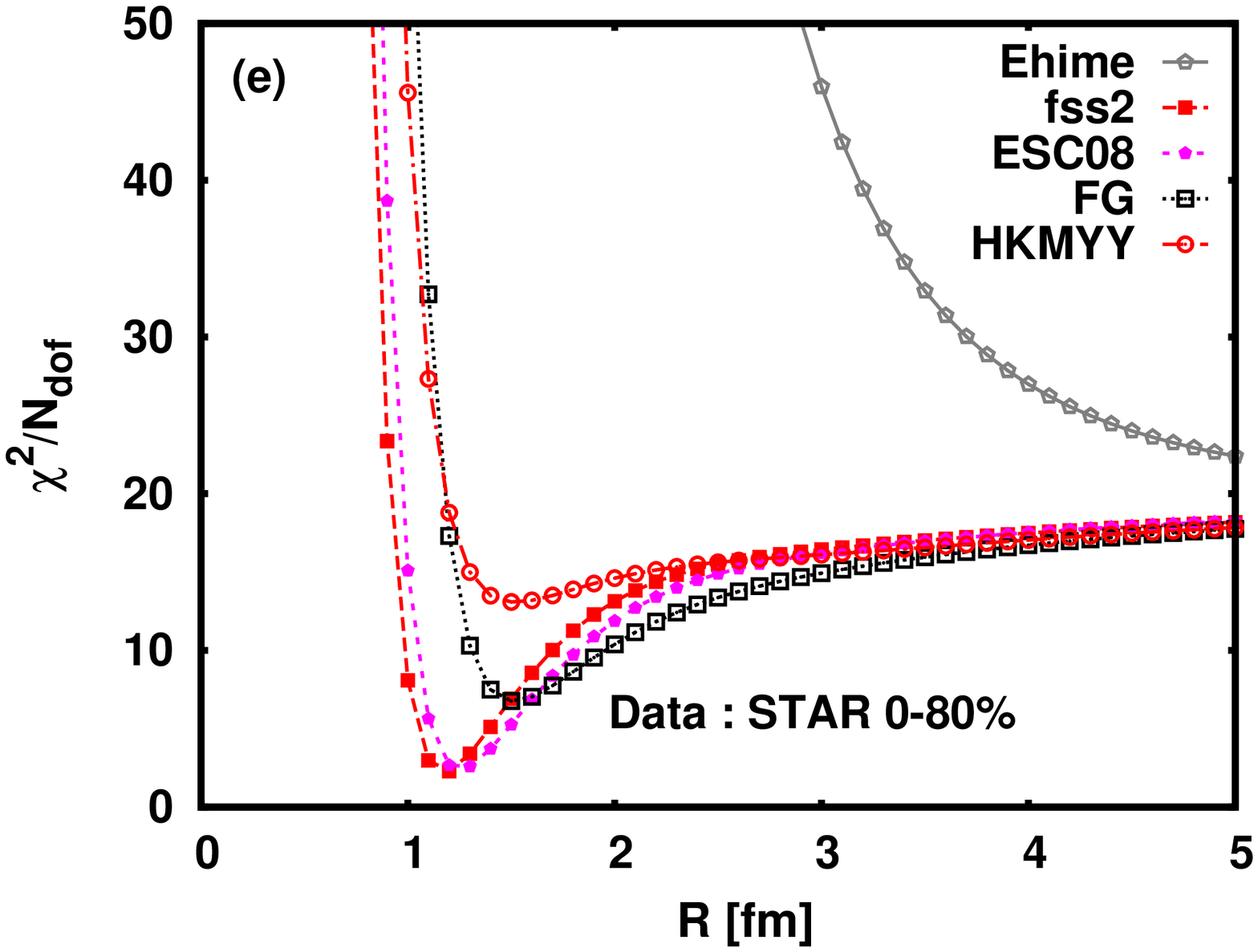}
  \caption{$\chi^2$ plotted against $R$ with \LL\ potentials.
  Nijmegen model D (ND, top left (a))
  Nijmegen model F (NF, top middle (b))
  Nijmegen soft core (NSC89, top right (c)),
  Nijmegen soft core (NSC97, bottom left (d)),
  and the rest of potentials in Tables \ref{tbl:potential}
  and \ref{tbl:potential_nagara} (bottom right (e)).}
  \label{fig:chi2stat}
 \end{figure*}

In relativistic heavy-ion collisions, the source emitting hadrons
shows collective behaviors. At the RHIC energy considered in this paper,
the hot medium produced in the collisions exhibits strongly correlated
property which is understood as nearly perfect fluid of the deconfined
quarks and gluons. The hadrons are produced at the hadronization from
the fluid, followed by rather dissipative transport processes
\cite{Gyulassy-McLerran}. 
As a result, the emission source $S(x,\boldsymbol{k})$ in Eq.~\eqref{eq:c2}
might not be characterized by a simple parametrization. 
In fact, the two-pion correlation functions
have been extensively discussed in heavy-ion collisions and 
are found to be sensitive not only to the emission source size
but to various aspects of the collision processes \cite{Pratt:HBTpuzzle}.
This fact made it difficult to fit the measured data 
within a simple model calculation even with collective effects being taken 
into account. 

We expect that the $\Lambda$ source may have a simpler form than
that for pions for the following reasons.
First, $\Lambda$ is expected to interact weakly with environments 
mainly consisting of pions. Single particle levels of $\Lambda$
including those of deep $s$-states are clearly observed~\cite{Lambda_Spectroscopy}.
This is in contrast with nucleons and pions, 
whose single particle states have large widths inside nuclei,
and suggests weaker interaction of $\Lambda$
with pionic environment, since nuclei contain many virtual pions.
Second, the decay feed effects are expected to be smaller.
It is known that the feed down effects of $\Lambda \to p\pi^-$
and $p\Lambda$ interactions are important to understand the $pp$
correlation function. 
For the \LL\ pair, there is no Coulomb suppression of low relative
momentum pairs
and contribution from particles which decay into $\Lambda$ is
limited.
In addition, we can in principle remove those $\Lambda$ particles from
weak decays such as $\Xi^- \to \Lambda \pi^-$ by using the vertex
detector.
In the following, we will show that data on $\Lambda\Lambda$ correlation
measured in heavy-ion collisions at the RHIC energy are useful to
discriminate the $\Lambda\Lambda$ interaction. 

To illustrate the capability, we first examine the correlation function
from a simple, static and spherically symmetric source
\begin{equation}
 S_{\text{stat}}(x,\boldsymbol{k}) = *\exp\left[-\frac{x^2+y^2+z^2}{2R^2}\right]\delta(t-t_0),\label{eq:sph_source}
\end{equation}
where normalization is omitted since it is canceled in the correlation
function. 

Putting Eqs.~\eqref{eq:wf} and \eqref{eq:sph_source} into
Eq.~\eqref{eq:c2} and projecting onto the function of
$Q=|\boldsymbol{Q}|$ by integrating out the angle variables, the
correlation function for the source function \eqref{eq:sph_source} becomes
\begin{align}
 C_{\text{stat}}(Q) &= 1-\frac{1}{2}e^{-Q^2 R^2} +
  \frac{1}{4\sqrt{\pi}R^3}\int_{0}^{\infty}dr \, r^2 e^{-r^2/4R^2}
 \nonumber\\
 &\times \left[ [\chi_Q(r)]^2-[j_0(Qr/2)]^2 \right].
\label{eq:c2_stat}
\end{align}
Thus the effect of interaction is incorporated as the difference of the
squared relative wave function. The free case is given by a simple
Gaussian, which is often used to obtain the source size.


In the static and spherically symmetric source model
\eqref{eq:sph_source}, the only parameter is the source size $R$.
Thus, it is convenient to calculate 
\begin{equation}
 \chi^2 \equiv \sum_{i} \left[
			 \frac{C(Q_i)^{\text{data}}-C(Q_i)^{\text{model}}}{\sigma_i^{\text{data}}} \right]^2
\end{equation}
as a function of $R$ then search for minimum in order to address which
potential is favored in data. 

In the following, we compare the model correlation function
\eqref{eq:c2_stat} with the data of $\Lambda\Lambda$ correlation
for $0.01 \leq Q_i < 0.5$~GeV,
measured by the STAR collaboration in Au+Au collisions at
$\sqrt{s_{NN}}=200$~GeV with 0-80\% centrality \cite{STAR}. 
$\chi^2$ is calculated for all the potential tabulated in Tables
\ref{tbl:potential} and \ref{tbl:potential_nagara} with the spherical
static source model $S_\text{stat}$.

%
%

The results are shown in Fig.~\ref{fig:chi2stat},
where $N_{\text{dof}}=24$.
One sees that the behavior of $\chi^2/N_{\text{dof}}$ as functions of
$R$ strongly depend on the choice of the potential. On the one hand,
some potentials exhibit monotonically decreasing behavior with $R$ then
asympotically becomes flat. These potentials typically have too strong
attraction (Ehime, ND50, NF46, ND46, NF44 etc) or
too large effective range (NSC97, NSC89).
On the other hand, a stable minimum with small $\chi^2/N_{\text{dof}}$
is achieved in the region $1~\mathrm{fm} < R < 1.5~\mathrm{fm}$
in potentials such as ND56, NF50, fss2 and ESC08.
Looking at the scattering length and the effective range of those potentials,
we find that there is a range of those quantities
in which the corresponding potentials show the stable 
and small $\chi^2$ minimum,
as shown in the shaded area in Fig.~\ref{fig:AR-sel}.
There are also marginal potentials (HKMYY, NSC89-820) which exhibit
also a stable minimum but with $\chi^2/N_{\text{dof}} > 5$.
Since we have compared with the raw data, 
these potentials may yield acceptable fit after appropriate corrections,
thus should not be ruled out.
NSC97d shows a different behavior from any other potentials,
due to the narrow effective range and small scattering
length as seen in Fig.~\ref{fig:AR-sel} which may not be realistic.
The small $\chi^2$ at $R=0.6$fm is achieved by approaching the long tail part
of $C(Q)$, rather than the interaction dominated part. 

 \begin{figure}[t]
  \includegraphics[width=3.375in]{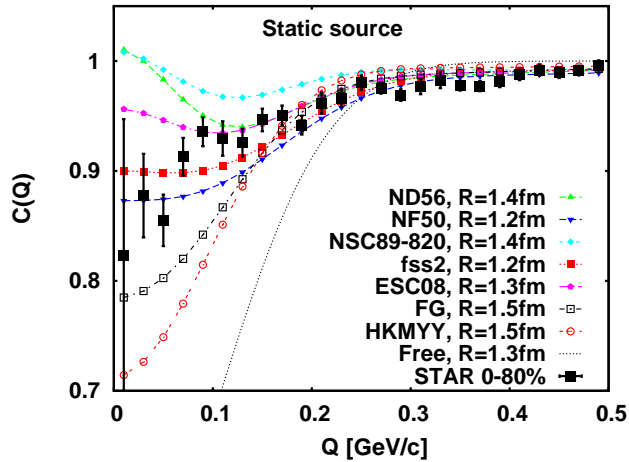}
  \caption{$\Lambda\Lambda$ correlation function from the static spherically
  symmetric source at the minimum of $\chi^2$ together with experimental
  data by STAR.}
  \label{fig:LL_correlation_static}
 \end{figure}

 \begin{figure*}[t]
  \includegraphics[width=3.375in]{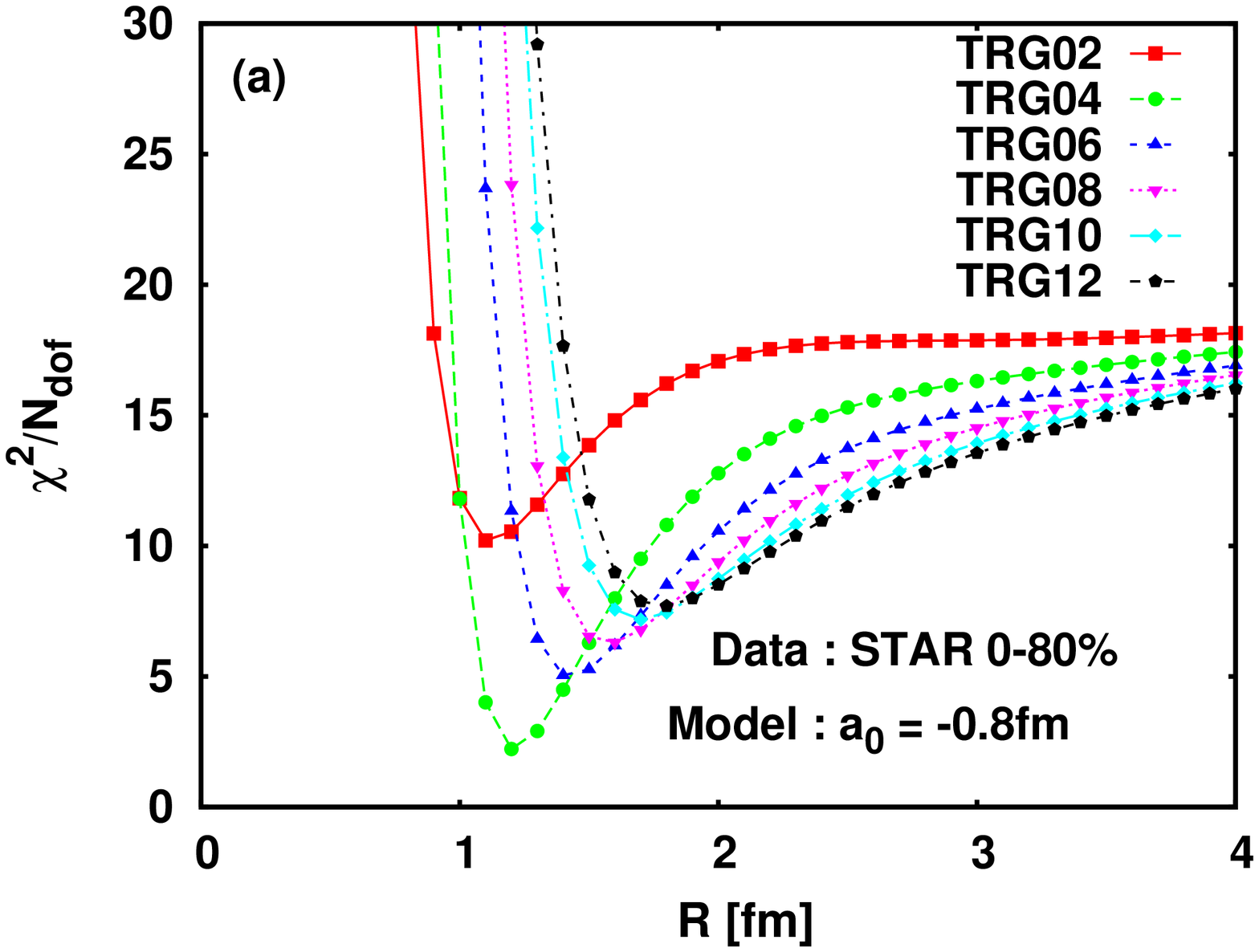}
  \includegraphics[width=3.375in]{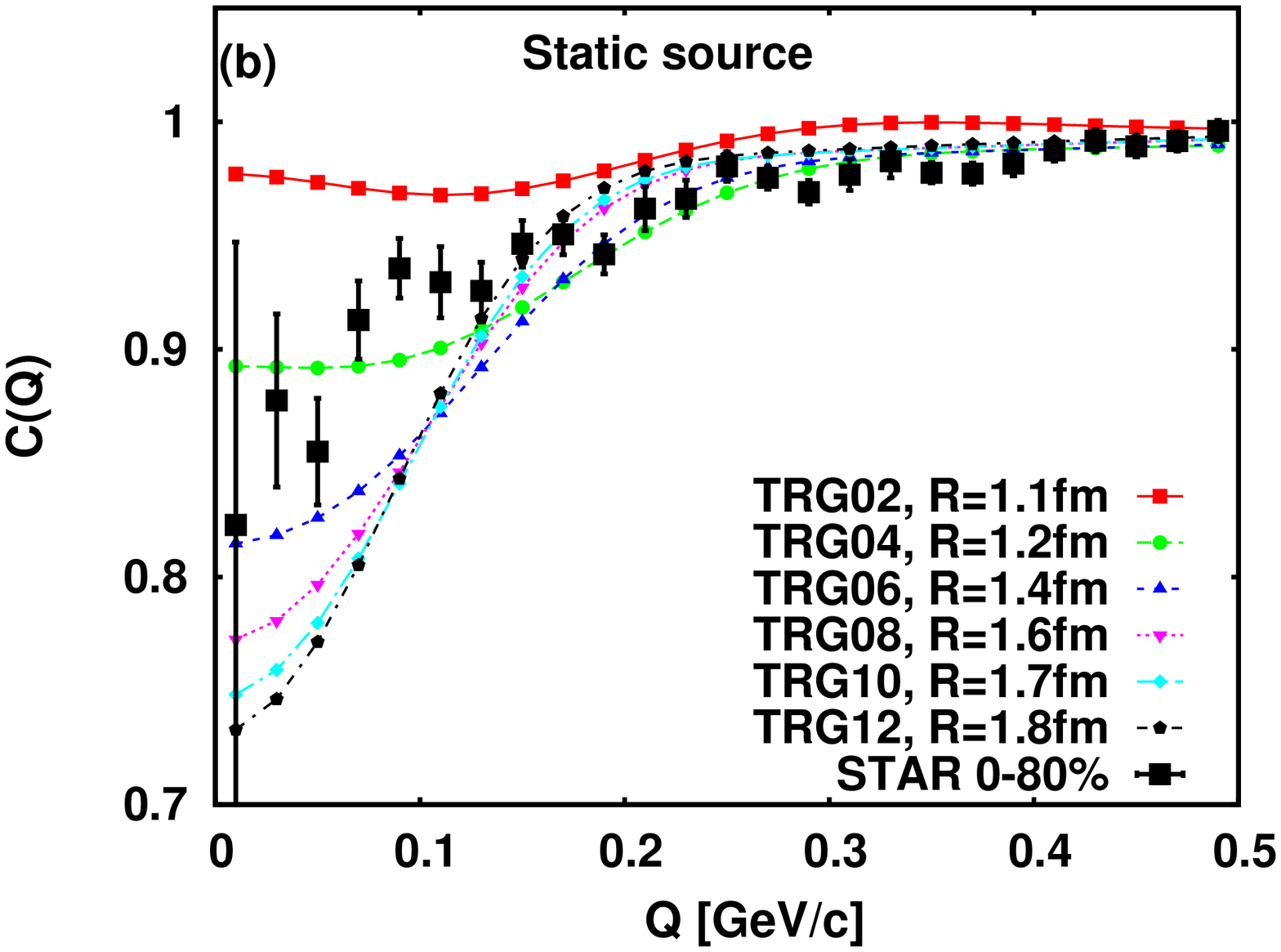}
  \caption{Results for TRG potentials. (a) $\chi^2/N_{\text{dof}}$
  against $R$. (b) $C(Q)$ at the minimum $\chi^2$.}
  \label{fig:chi2_TRG}
 \end{figure*}

Figure \ref{fig:LL_correlation_static} displays
the correlation functions for potentials considered here
compared with the experimental data.
The size parameter $R$ adopted in the figure corresponds to the minimum of
$\chi^2$ in Fig.~\ref{fig:chi2stat}.
We also plot the free correlation function, Eq.~\eqref{eq:c2_stat}
without the last term, for comparison.
While the values of $\chi^2$ do not differ much, the correlation
function at low $Q$ shows a substantial variation among the potentials.
The small difference of $\chi^2$ is attributed to larger error bars in
the experimental data at low $Q$ region.
One sees that results from these potentials have all $C(Q) > 1/2$ at low
$Q$, fairly reflecting the attraction between two $\Lambda$. 
Among the potentials with small minimum $\chi^2$,
on the one hand, NF50 and fss2 show $C(Q < 0.1~\GeV)\sim 0.9$ and give a
good description for the tail part around $Q \sim 0.2~\GeV$.
On the other hand, ND56 and ESC08 exhibit an weak enhancement
at $Q < 0.1~\GeV$ with a good fit to the data at  $0.1~\GeV < Q < 0.3~\GeV$. 
Therefore, precision measurement at $Q < 0.1$ GeV will provide
further constraints on the $\Lambda \Lambda$ interaction.
According to Table \ref{fig:AR-sel}, those potentials have 
$-1.2~\fm < a_0 < -0.8~\fm$ and $3.2~\fm< r_{\text{eff}}< 6.5~\fm$.
However,
the effective range is not contrained well because 
the model potentials do not have a combinations of a large scattering
length and a large effective range.
Thus, we further construct model potentials
by varying the effective range with a fixed scattering length
$a_0 = -0.8$fm in the two-range Gaussian (TRG) form
\eqref{eq:TRG}. Results for the $\chi^2/N_{\text{dof}}$ and
corresponding $C(Q)$ are displayed in Fig.~\ref{fig:chi2_TRG}.
One sees that there always exists
a minimum in the $\chi^2$ plot which is
particularly sensitive to variation of the effective range around
$r_{\text{eff}}\simeq 4~\fm$. The global minimum achived for
$r_{\text{eff}}\simeq 4~\fm$ is in accordance with the above model
analysis. Since the behavior of $C(Q)$ with large effective range,
$r_{\text{eff}} > 6$ fm is similar to FG and HKMYY in
Fig.~\ref{fig:LL_correlation_static}, a large effective range might be
favored if the data receive the correction. Consequently, the present
analysis provides rather limited constraints on the effective range than
the scattering length. The favored range of the scattering length and
the effective range is indicated by the shaded area in Fig.~\ref{fig:AR-sel}.

\begin{table}
\caption{$\Lambda\Lambda$ potential parameters with various effective ranges for
 fixed scattering length, $a_0=-0.8$fm in the two-range Gaussian (TRG)
 form \eqref{eq:TRG}.}
\label{tbl:TRGpotential}
\begin{tabular}{l|r|rrrr}
\hline
\hline
Model & $r_\mathrm{eff}$ (fm) & $\mu_1$ (fm) & $V_1$ (MeV)
& $\mu_2$ (fm) & $V_2$ (MeV)
\\
\hline
TRG02	& $2.0$  & 0.6 &  $-405.97$  & 0.45 & 582.29 \\
TRG04	& $4.0$  & 0.6 &  $-1835.95$ & 0.45 & 4976.17\\
TRG06	& $6.0$  & 0.6 &  $-5569.58$ & 0.45 & 25435.95\\
TRG08   & $8.0$  & 0.8 &  $-889.42$  & 0.45 & 14595.38\\
TRG10   & $10.0$ & 0.8 &  $-1440.02$ & 0.45 & 43254.42\\
TRG12   & $12.0$ & 1.0 &  $-358.38$  & 0.45 & 20522.96\\
\hline
\end{tabular}
\end{table}

\subsection{Wave functions}

In order to characterize the potentials which give reasonable
description of the measured $\Lambda\Lambda$ correlation data, 
we discuss the corresponding potentials and resultant wave functions. 

 \begin{figure*}[t]
  \includegraphics[width=3.375in]{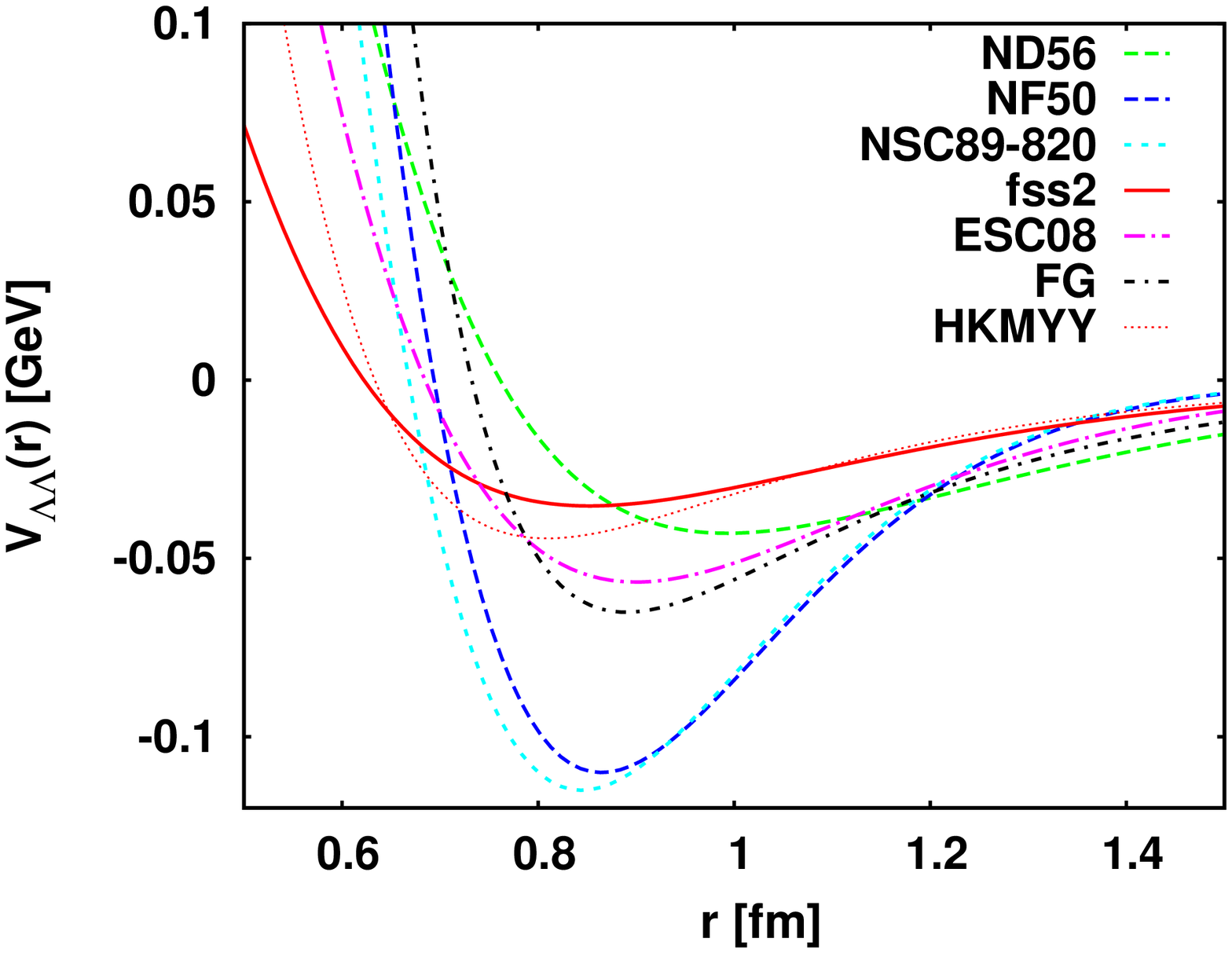}
  \includegraphics[width=3.375in]{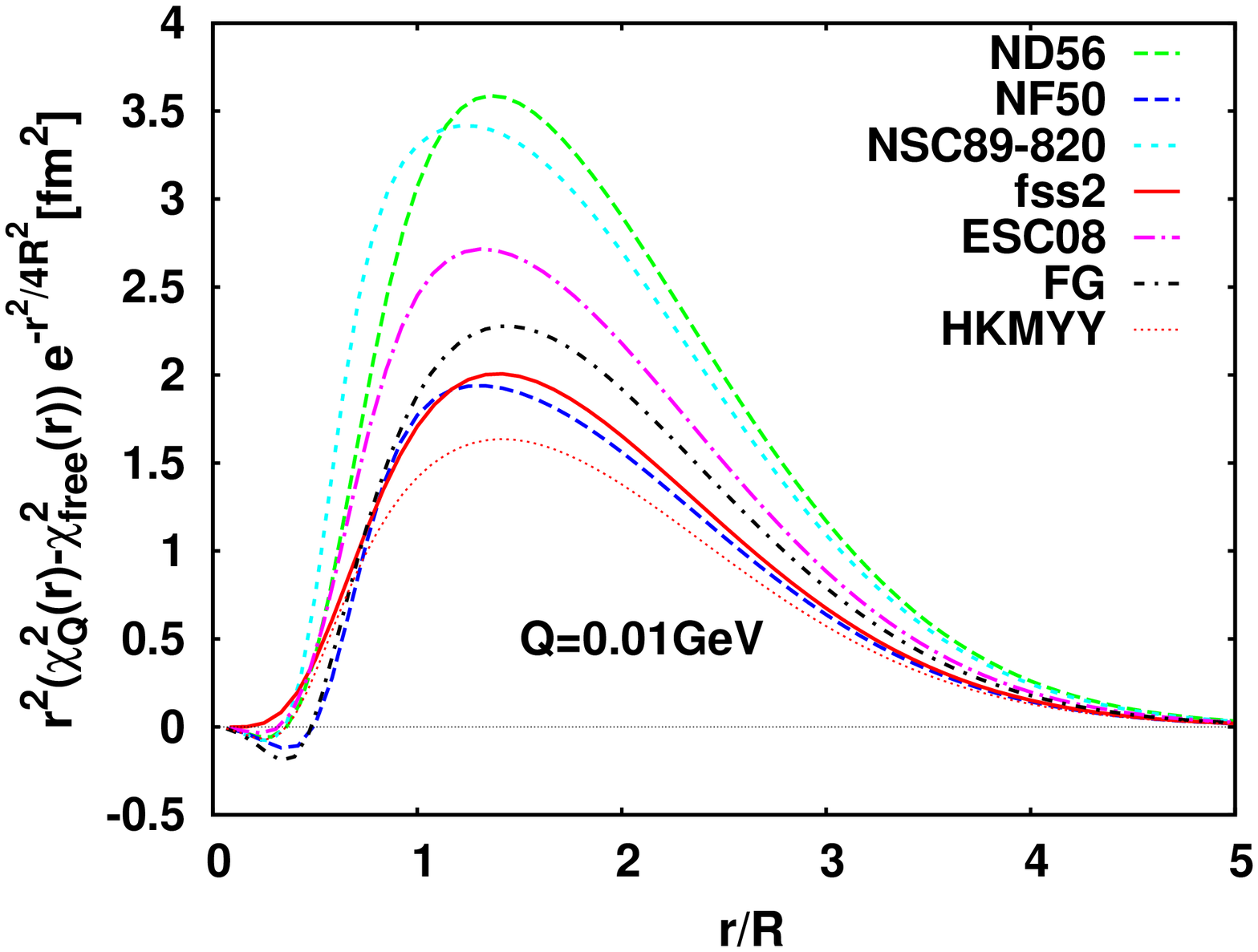}
  \caption{Left: $\Lambda\Lambda$ potential for the selected parameter sets
  which fit the measured data well. Right: Deviation of the relative wave functions from the free one weighted with the source function at $Q=0.01$GeV.}
  \label{fig:wf}
 \end{figure*}

Since effects of interaction on the correlation function are
incorporated through the difference from the free wave function as seen
in Eq.~\eqref{eq:wf}, we display the integrand of the last term of
Eq.~\eqref{eq:c2_stat} rather than the wave function itself in the right
panel of Fig.~\ref{fig:wf} as well as the potentials in the left
panel. The horizontal axis in the right panel is normalized by the size
parameter $R$ at the minimum $\chi^2$ to reduce the apparent effect due to the
different size in the source function. Although the wave function
$\chi_Q(r)$ reflects the remarkable differences of the potentials at
small $r$, $r^2e^{-r^2/4R^2}$ acts as a
weight factor such that the behavior of the correlation function is most
sensitive to the wave function at $r \simeq 1-3$ fm.
One sees that the deviation from the free wave
function is fairly reflected onto $C(Q)$.  Namely, ND56 and NSC89-820
which show the largest attraction in the wave function at $r/R \simeq 1.5$ exhibit the strongest bunching in $C(Q)$, as
seen in Fig.~\ref{fig:LL_correlation_static}. ESC08, NF50 and fss2
also follow this trend. HKMYY, one of the two wave functions motivated by the Nagara
event, has the weakest attraction among the potentials thus leads to the
smallest deviation from $C(Q=0)=0.5$ in
Fig.~\ref{fig:LL_correlation_static}. The other one, FG, has a
somewhat stronger attraction, but the strongest repulsion around the
origin leads to $\chi^2_Q < \chi^2_{\text{free}}$ at $r < 0.5~\fm$, which
finally gives $C(Q)\simeq 0.8$. It is instructive to note that 
fss2 and NF50 give a similar wave function despite the difference between the
potentials. As given in Table \ref{tbl:potential}, both effective range
and scattering length have close values in these potentials. This
indicates that  the $C(Q)$ is essentially determined by $a_0$ and
$r_{\text{eff}}$ rather than the detailed form of the potential. 

The above consideration demonstrates how sensitive the correlation
function $C(Q)$ is to the relative wave function. From the analysis with
the static spherically symmetric source, strong constraints on the
effective range and the scattering length of $\Lambda\Lambda$
interaction have been obtained.
The next step is to
examine whether this capability is affected by the dynamics of 
relativistic heavy-ion collisions at the RHIC energy. 

\section{Effect of collectivity}
\label{sec:flow}
In the following, we consider source functions incorporating effects of
expansion dynamics of the relativistic heavy-ion collisions. 
We assume that $\Lambda$ particles are produced at the
chemical freeze-out following the hadronization from quark-gluon plasma.
Although $\Lambda$ may interact with other produced hadrons and
substantial amount of $\Lambda$ multiplicity comes from decay of heavier
particles, we assume that the information on $\Lambda\Lambda$
interaction encoded in the correlation function is not distorted much
by those hadronic effects.
This simplification may be justified in part by the smaller
$\Lambda$ interaction with the pion dominant environment
and smaller feed down effects expected in \LL\ correlation.
Especially, if most of the weak and electromagnetic decay 
to $\Lambda$ is removed by using the vertex detector,
feed down effects on $\Lambda$ is expected to be small.
We examine the feed down effects in the next section and
concentrate on the effects of collectivity in this section.

\subsection{Models}

Effects of the collective flow can be studied by modifying
the source function $S(x,\boldsymbol{k})$. We utilize a thermal source 
model to implement the collective effect and change the source
geometry to more relevant one for relativistic heavy-ion collisions. The
collective flow velocity at the emission point $x$ is denoted by $u^\mu$
and the Fermi distribution function $n_F(E,T)=(e^{E/T}+1)^{-1}$ for
$\Lambda$ is introduced to accommodate the thermal distribution in the
rest frame of the emission point. 

The source function is generally defined through the invariant spectrum
\begin{equation}
 E\frac{d^3N}{d\boldsymbol{k}^3} = \int d^4x S(x,\boldsymbol{k}).
\end{equation}
We consider several source functions to discriminate effects
on the correlation function. As the simplest extension from the static
source model used in the previous section, we implement the collective flow as
\begin{align}
 S_{\text{sph}}(x,\boldsymbol{k}) &= * u\cdot k \, n_F(u\cdot k, T) \nonumber\\
 &\times \exp\left[ - \frac{x^2+y^2+z^2}{2R^2}\right]\delta(t-t_0).\label{eq:sphflow_source}
\end{align}
The spherically symmetric flow velocity,
$u^\mu = \gamma (1,\boldsymbol{v})$ where 
$\gamma = (1-v^2)^{-1/2}$ , is assumed to exhibit a Hubble-type
expansion  $v = \tanh (\eta_r r/R)$. The coefficient
$\eta_r$ controls the strength of the expansion. The product 
$u \cdot k$ brings the space-momentum correlation into the source
function if $\boldsymbol{v}\neq 0$.
When $\eta_r=0$, the space-momentum correlation of the source function
is lost, i.e., the source function is factorized into
$S(x,\boldsymbol{k})= A(x)B(\boldsymbol{k})$ then the resultant
correlation function reduces to that of the static source 
\eqref{eq:c2_stat} because the momentum dependent part
$B(\boldsymbol{k})$ is canceled by the denominator in Eq.~\eqref{eq:c2}.

While the above model \eqref{eq:sphflow_source} is useful to understand
effects of the collective flow, the geometric part of the source
function still posses the spherical symmetry which is not appropriate for
heavy-ion collisions at the RHIC energy. Owing to the huge colliding
energy, the system undergoes rapid expansion along the collision axis
(we take it as $z$ axis) followed by slower one in the perpendicular
direction. 
Therefore, we consider a cylindrically symmetric expanding system with
the longitudinal boost invariance.
The boost invariance also implies the infinite
extent of the source in the collision axis. 
In the presence of the strong longitudinal flow, however, this does not
mean that source function have infinite width in the longitudinal direction
since the thermal factor naturally gives a finite extent. In the Boltzmann
approximation, one can derive an approximate but analytic expression for the
longitudinal source size, $R_L \simeq \tau_0 \sqrt{T/m_T}$ \cite{Sinyukov},
where $\tau_0$ denotes the freeze-out proper time
and $m_T$ is the transverse mass $m_T = \sqrt{p_x^2+p_y^2+m^2}$.
The boost invariant approximation is of
course valid only around midrapidity region. At the RHIC energy, effects
of possible deviation were found to be small \cite{morita_hydro}.
 
In reality, collisions have finite impact parameters
even in the highest multiplicity bin and event-by-event fluctuations may induce
further asymmetry to the source. Those effects may be important for
understanding experimental data on the $\pi\pi$ HBT correlation, but we
shall ignore them since we are not aiming at extracting the source
geometry but examining the final state interaction between
$\Lambda$. 
The present statistics does not seem enough to project
$\Lambda \Lambda$ correlation function onto each directions of
$\boldsymbol{Q}$. Thus, we assume the azimuthally symmetric gaussian
source profile. 
We expect this only leads to a smaller effective source size
resulting from averaging over the azimuthal angle. In this case, we can
put the average momentum $\boldsymbol{K}$ as
$\boldsymbol{K}=(K_T,0,K_z)$ without loss of generality.
Then a convenient source function has been used for the analysis of pion HBT radii
in Ref.~\cite{Sourcemodel},
\begin{align}
 S_{\text{cyl}}(x,\boldsymbol{k}) &= \frac{m_T\cosh(y-Y_L)}{(2\pi)^3\sqrt{2\pi(\Delta
 \tau)^2}} n_f(u\cdot k,T)  \nonumber \\
 &\times \exp\left[-\frac{(\tau-\tau_0)}{2(\Delta \tau)^2} -
 \frac{x^2+y^2}{2R^2}\right]\label{eq:source}.
\end{align}
where $y=1/2 \log[(E_k+k_z)/(E_k-k_z)]$ is the rapidity of the emitted
particle. The gaussian form of temporal part of the source function in
the proper time $\tau=\sqrt{t^2-z^2}$ takes into account possible
emission duration $\Delta \tau$ around a freeze-out time $\tau_0$. 

The four flow velocity can be parametrized by the longitudinal and
transverse rapidity
\begin{align}
 u^t&= \cosh Y_T \cosh Y_L\\
 u^z&= \cosh Y_T \sinh Y_L \\
 u^x&= \sinh Y_T \cos\phi\\
 u^y&= \sinh Y_T \sin\phi .
\end{align}
The longitudinal flow rapidity is given by the scaling solution \cite{Bjorken}
\begin{equation}
 Y_L = \eta_s = \frac{1}{2}\ln \frac{t+z}{t-z}
\end{equation}
The transverse flow rapidity is assumed to be
\begin{equation}
 Y_T = \eta_f \frac{r_T}{R}
\end{equation}
where $r_T = \sqrt{x^2+y^2}$ and $\eta_f$ controls the strength of the
transverse flow. 

We fix the transverse flow strength parameter
$\eta_f$ from the single $\Lambda$ spectra in 200GeV Au+Au collisions
measured by the STAR collaboration
\cite{agakishiev12:_stran_enhan_in_cu_cu}.
We use the $20-40$\% data since the detailed spectrum after
feed-down subtraction is given. 
As we have ignored the baryon and the strangeness chemical potentials in
the thermal description, we average the $\Lambda$ and $\bar{\Lambda}$
data for the fit. Fixing $T = 160$MeV, for the chemical freeze-out, 
we obtain the minimum of $\chi^2/N_{\text{dof}}$ is about 1.9 for
$\eta_f=0.33$. 
We use this value in the calculations below.\footnote{We checked that
the choice of the temperature and the resultant flow strength do not
influence our main objective in this paper, by repeating the same
calculations for $T=200$MeV and 120MeV. }
Note that this can be determined independent of the size
parameters, $R,\tau_0$, and $\Delta\tau$. These parameters are
regarded as free parameters to study the influence of the source
geometry on $\Lambda\Lambda$ correlation function. Relations to the HBT
radii have been extensively studied in Refs.~\cite{Sourcemodel} and
\cite{morita_hydroHBT} in the context of the $\pi\pi$ correlation.

\subsection{Effects of collectivity}
\label{sec:hbt}

Now we calculate the $\Lambda\Lambda$ correlation function for the expanding
sources. Owing the space-time correlation induced by the collective
flow, the correlation function depends on the average momentum
$\boldsymbol{K}$ in addition to the relative one. We integrate the
numerator and the denominator of Eq.~\eqref{eq:c2} as
\begin{equation}
 C(Q) = \frac{\int d^3 \boldsymbol{K}
  W_2(\boldsymbol{k_1},\boldsymbol{k_2})}{\int d^3 \boldsymbol{K} W_1(\boldsymbol{k_1})W_1(\boldsymbol{k_2})}.\label{eq:integrated_correlation}
\end{equation}
For the cylindrical source \eqref{eq:source}, the integration is carried
out for the rapidity and the transverse momentum within the experimental
acceptance ranges. For the spherical source, integration with respect to
$|\boldsymbol{K}|$ is done for the same range as the transverse momentum
in the cylindrical case, for simplicity. Although the following results
may change quantitatively due to the average momentum dependence of the
correlation function, these integrations do not lead to any change in
our discussion below. 

 \begin{figure}[ht]
  \includegraphics[width=3.375in]{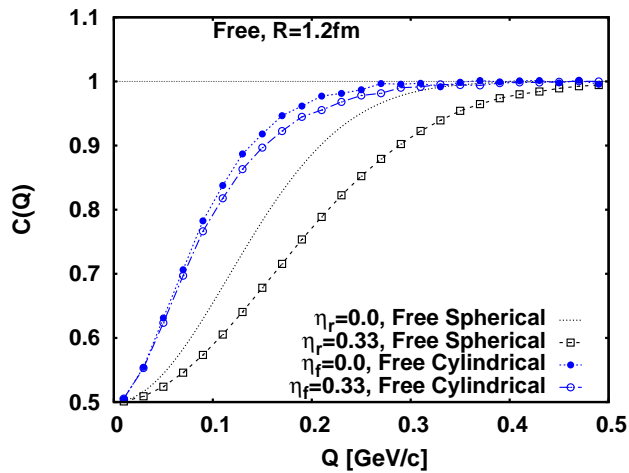}
  \caption{Free correlation functions for the spherically symmetric
  source \eqref{eq:sphflow_source}
  and for the cylindrically symmetric boost-invariant source
  \eqref{eq:source}. Closed and open symbols stand for $C(Q)$ with and
  without flow, respectively.}
  \label{fig:c2_free_flow}
\end{figure}

\begin{figure}[ht]
  \includegraphics[width=3.375in]{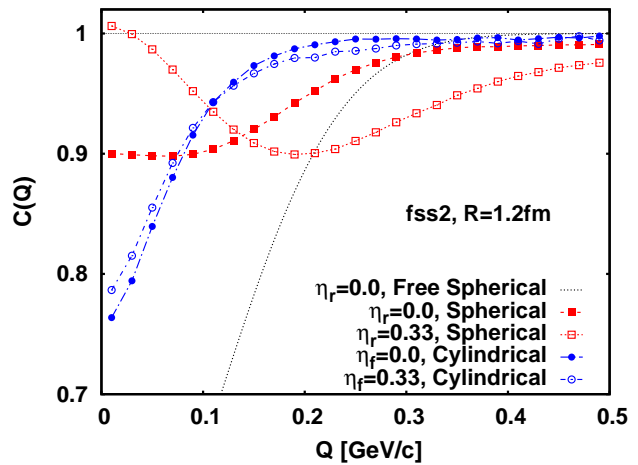}\\
  \caption{Correlation functions with fss2 interaction. Symbols are
  similar to the previous figure \ref{fig:c2_free_flow}. }
  \label{fig:c2_fss2_flow}
 \end{figure}

We begin with examining effects of collective flow in the spherically symmetric
source to make a clear connection with the analyses in the previous
section. We choose $R=1.2~\fm$ in the model source function
\eqref{eq:sphflow_source}. Although not fitted to the experimental data, we
apply the result of the fit to $p_T$ spectrum in the cylindrical source
model and put $\eta_r = 0.33$. Figure \ref{fig:c2_free_flow} illustrates
the effect of the flow strength on the free correlation functions.
Dotted line stands for the static case $1-\frac{1}{2}e^{-Q^2R^2}$.
The expanding case is represented by open squares. One sees that the flow makes
the correlation extended in the higher momentum region. In other words, 
the effective source size becomes smaller, as is well known for pion HBT radii
\cite{Pratt,Schlei}. In Fig.~\ref{fig:c2_fss2_flow}, we turn on the
interaction between $\Lambda$ by taking the fss2 potential.
Since the difference of the potentials results in the relative wave function,
our discussion in this section does not depend on the choice of the
potential. By comparing with the $\eta_r=0$ case, which is the same as fss2
result in Fig.~\ref{fig:LL_correlation_static}, one finds that
the correlation function resembles the more attractive potential. Since
the interaction is not altered, one can understand this behavior as the
effect of the collective flow. As shown in Fig.~\ref{fig:c2_free_flow},
effective source size is decreased by the collective flow. This can be
understood as the result of position-dependent suppression of emission
probability by the thermal factor \cite{morita_hydroHBT}. This makes
$\Lambda$ pairs with small $Q$ concentrated inside small region thus
they become more sensitive to the short range interaction. 

Results for the cylindrical source function $S_{\text{cyl}}$
\eqref{eq:source} are also displayed in
Figs.~\ref{fig:c2_free_flow} and \ref{fig:c2_fss2_flow}.
Here we put $\tau_0= 5$fm and $\Delta \tau =2$fm.
The free case with $\eta_f = 0$ is shown as closed circles in
Fig.~\ref{fig:c2_free_flow}. The narrower width of the correlation
function than that of the spherical source is due to the fact that $\eta_f$
controls the transverse flow only and that the boost-invariant longitudinal expansion 
takes place. As seen in Eq.~\eqref{eq:source},
the boost-invariant source function has infinite extent in the
longitudinal direcion if one ignores the thermal distribution. Although
the longitudinal boost invariant expansion makes the effective source
size finite, the source function has still a large longitudinal
extent such that the behavior of $C(Q)$ resembles larger source size than
the spherically symmetric source.
Since the longitudinal flow effect is dominant, the effect of the
transverse flow seen through $\eta_f$ is small. Note,
however, that the effective source size becomes smaller by increasing
$\eta_f$, because the same discussion as the case of the spherical source
applies to the present result. As a result, the whole shape of the $C(Q)$
does not change much by the interaction.  Nevertheless, the behavior of
$C(Q)$ at low $Q$ remains sensitive to the interaction, as depicted in
Fig.~\ref{fig:c2_fss2_flow}. If we
introduce a finite longitudinal size into the source function, such as
$e^{-z^2/2R^2}$, the behavior of the
correlation function becomes closer to the case of the spherical source
since the effective source size is reduced. We note that increasing
$\tau_0$ as well as $\Delta \tau$ also makes the effective source size
larger, since $\tau_0$ corresponds to the source extent in $z$ direction
and $\Delta \tau$ contributes to the correlation function through the
emission time difference in Eq.~\eqref{eq:c2}.

 \begin{figure}[ht]
  \includegraphics[width=3.375in]{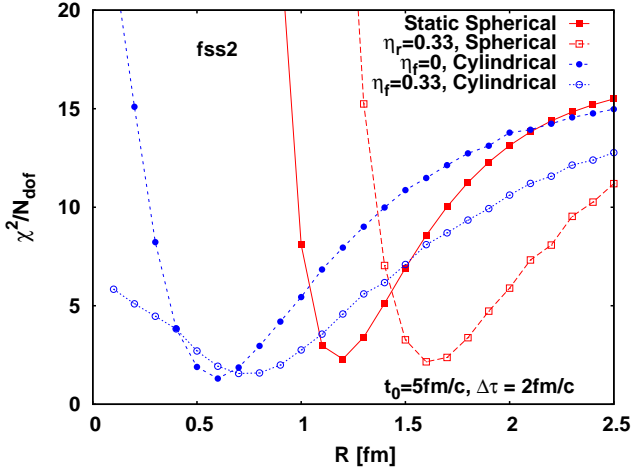}%
  \caption{$\chi^2/N_{\text{dof}}$ against the size parameter for the
  fss2 interaction. }
  \label{fig:chi2_fss2}
 \end{figure}

\begin{figure}[ht]
  \includegraphics[width=3.375in]{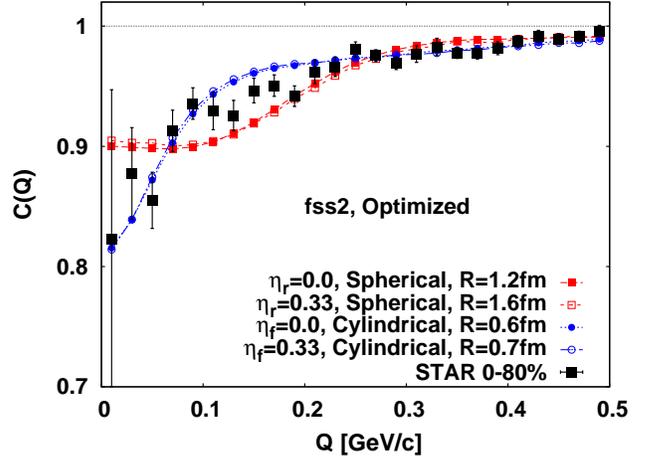}
  \caption{$\Lambda\Lambda$ correlation function with fss2 interaction
  for the spherical and the cylindrical boost-invariant source optimized
  for minimum $\chi^2$.}
  \label{fig:c2_fss2_cyli}
 \end{figure}

As a result of geometry and flow effects,
optimal source size is modified from that in the spherical static source case.
In Fig.~\ref{fig:chi2_fss2},
we show the size parameter dependence of $\chi^2/N_\mathrm{dof}$
with fss2 interaction.
For the spherical geometry,
the optimal source size is shifted to the larger direction by 30-40 \%.
Figure \ref{fig:c2_fss2_cyli} shows the corresponding correlation
functions at the minimum of $\chi^2$. Interestingly, despite the
difference in the source size, the resultant correlation functions for
$\eta_r=0$ and $0.33$ with the spherical source are almost the same.
The effect of the flow is absorbed into the larger optimal source size.

For the cylindrical and boost invariant source,
the optimal transverse source size is also shifted upwards,
while the shift is smaller as discussed above. The behavior of the
$\chi^2$ in small $R$ reflects the effect of the geometry and the flow. 
Since the source function is elongated in the longitudinal direction,
the source has a longitudinal extent even when $R$ is so small.
Thus $\Lambda$ particles feel less attraction than in the case of the
spherical source with small $R$. The transverse flow further reduces 
the variation of $C(Q)$ against $R$. It is interesting to note that
despite almost the same value of $\chi^2/N_{\text{dof}}$ of the
cylindrical source as that of the spherical source, the optimized
correlation function shown in Fig.~\ref{fig:c2_fss2_cyli} has a
different feature. The result of the cylindrical source indicates that
$C(Q)$ with the small width also describes the data well if this
shape of $C(Q)$ can be obtained, though it was not possible  in the spherical source as shown
in Sec.\ref{sec:source}.

Then, we have performed the same analysis for other \LL\ interactions.
We found that the favored \LL\ interactions are the same
as those in the case of the spherical static source $S_{\text{stat}}$.
However, we also found that interactions with larger effective range and
smaller scattering length are favored. For example, among ND
interactions, ND58 is now better than ND56, contrary to the
case of $S_{\text{stat}}$ (Fig.~\ref{fig:chi2stat}),
while ND56 still
gives a good fit, too. NSC97 potentials are also found to be improved,
but $\chi^2/N_{\text{dof}} \simeq 8$ at the best in NSC97c and NSC97e.
NSC97d no longer shows a good fit with the cylindrical source
model $S_{\text{cyl}}$. This indicates that the good fit in
$S_{\text{stat}}$ (Fig.~\ref{fig:chi2stat}) 
was spurious one;
not due to the interaction but to an accidental geometric effect. 

 \begin{figure*}[ht]
  \includegraphics[width=3.375in]{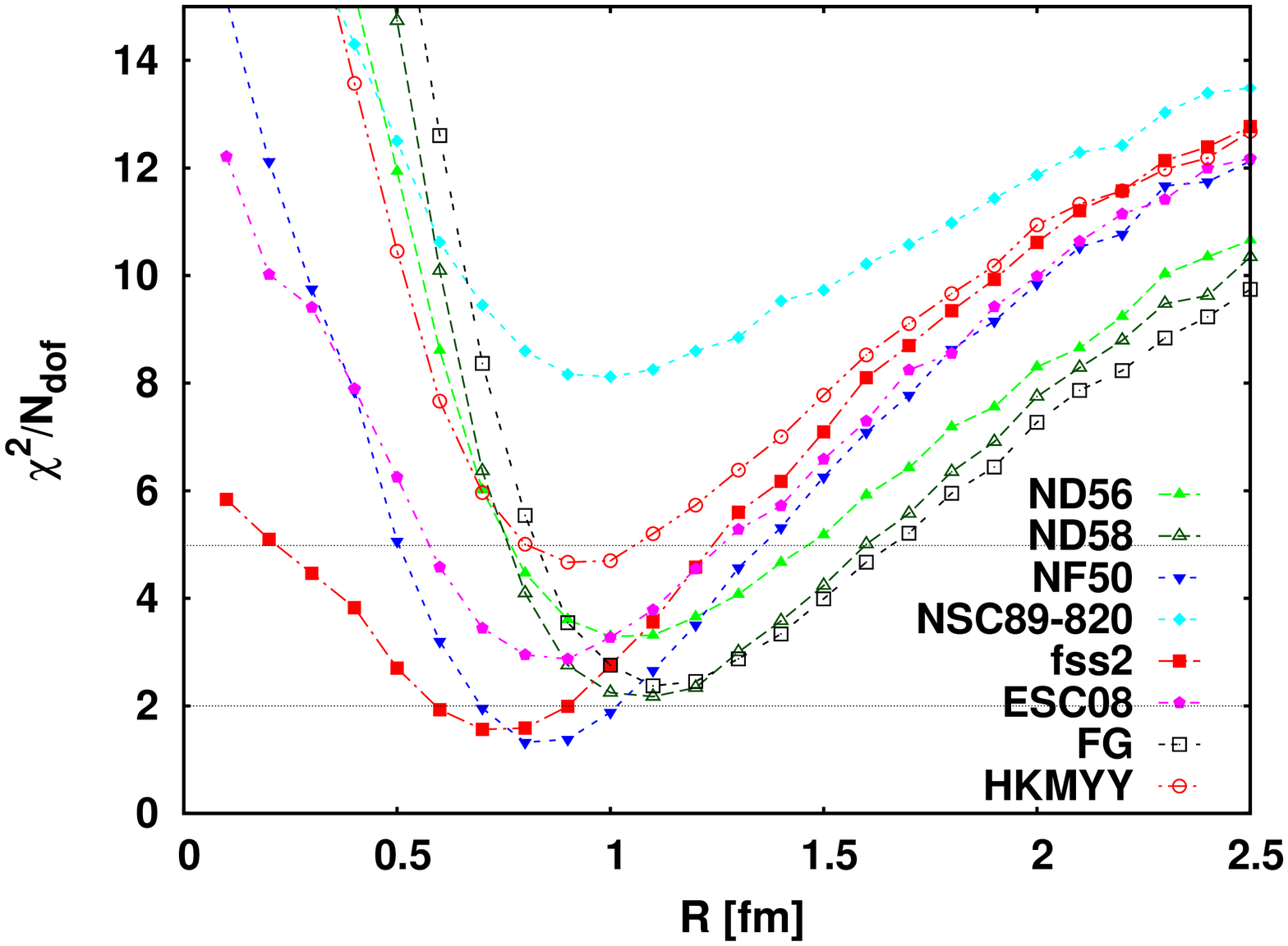}
  \includegraphics[width=3.375in]{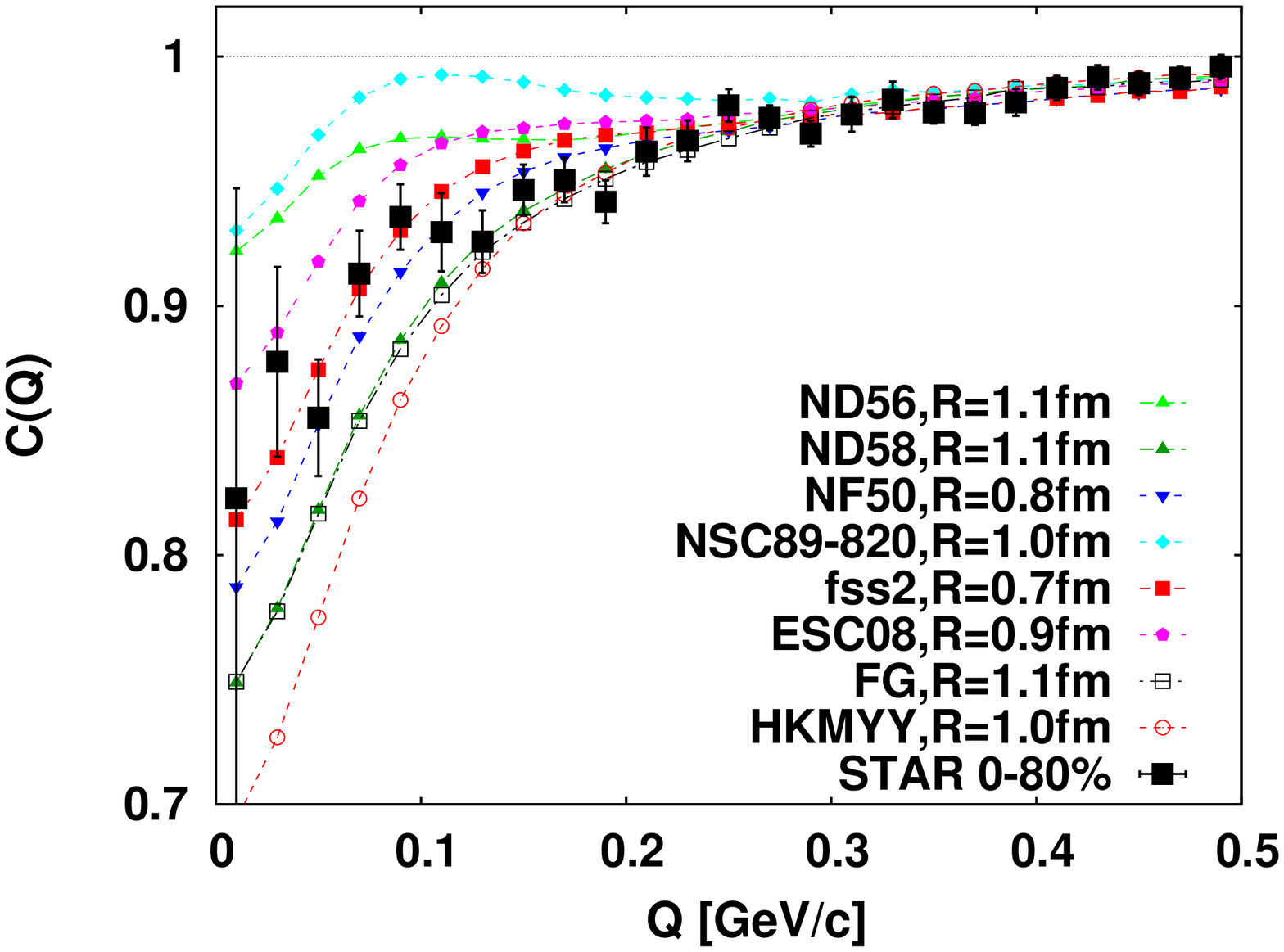}
  \caption{Left: $\chi^2/N_{\text{dof}}$ for the favored potentials in
  the cylindrical source model with collective flow
  \eqref{eq:source}. Right: corresponding $\Lambda\Lambda$ correlation
  functions at the minimum $\chi^2$.}
  \label{fig:cyli}
 \end{figure*}

$\chi^2$ and the optimized $C(Q)$ for all the potential that give
$\chi^2/N_{\text{dof}} < 8$ are summarized in Fig.\ref{fig:cyli}.
Since ND58 and FG have almost same $a_0$ and $r_{\text{eff}}$, the
result is so. 
Other \LL\ interactions give larger $\chi^2$.
In Fig.~\ref{fig:AR-sel}, we mark the favored \LL\ interation with big circles.
The scattering parameters of these interactions are in the range
$-1.8~\mathrm{fm}^{-1} < 1/a_0 < -0.8~\mathrm{fm}^{-1}$ and 
$3.5~\mathrm{fm} < r_\mathrm{eff} < 7~\mathrm{fm}$.
Potentials studied in this work and outside of this region
give larger $\chi^2$, $\chi^2/N_\mathrm{dof} > 5$,
at any value of the source size and with any geometry and flow values.
One sees that the small size is favored owing to somewhat scattered data
points and larger errors in small $Q$ region. The difference of the
correlation function in those potentials are of the same size as
experimental errors for $0.1 < Q < 0.2$ GeV and even larger at 
$Q < 0.1$ GeV. Therefore, the data give
strong constraint on the interaction potential of the $\Lambda \Lambda$
system. Indeed, the favored potentials obtained from this analysis
agrees well with 
the results obtained in the analysis of 
$^{12}\mathrm{C}(K^-,K^+)\Lambda\Lambda X$ reaction~\cite{E522}.
Enhancement of the \LL\ invariant mass spectrum at low energies
are well described by the final state interaction effects
with fss2~\cite{fss2} and ESC04d~\cite{ESC04}.
The scattering length and the effective range in ESC04d interaction
are $a_0=-1.323~\fm$ and $\reff=4.401~\fm$, respectively,
which are inside the region obtained in the present analysis.
The analyses of $\DBLL$ in the Nagara event~\cite{FG,Hiyama}
show \LL\ interactions which are inside the allowed region.

\section{Feed-down contribution}
\label{sec:feeddown}

\subsection{Estimate of feed-down contribution}
So far we have assumed that $\Lambda$ baryons are directly
emitted from the hot matter. This assumption would be valid if one could
remove decay contribution from parent particles such as $\Xi$ and
$\Sigma^0$ or the correlation function $C(Q)$ was not affected by such
feed-down contribution. In the case of the $\pi\pi$ correlation
function, it has been known that long-lived parents give a sharp
correlation  near $Q\simeq 0$ which cannot be resolved 
thus cause an apparent reduction of the
intercept $C(Q=0)$ \cite{Grassberger1977,Csorgo1996,Wiedemann-Heinz}.
The same argument  applies to the $\Lambda\Lambda$ correlation as well.
For $N_{\text{tot}}$ being  the total number of  measured $\Lambda$
and $N^p$ being the long-lived parents decaying into $\Lambda$, respectively,
the effective intercept $\lambda$ is given by
\begin{equation}
 \lambda = \left( 1- \frac{N^p}{N_{\text{tot}}} \right)^2.
\end{equation}
The correlation function after the
feed-down correction to the bare one reads
\begin{equation}
 C_{\text{corr}}(Q) = 1+\lambda(C_{\text{bare}}(Q)-1),
\label{Eq:corrbare}
\end{equation}
which should be confronted with the data.

The contribution to $N_{\text{tot}}$ mainly consists of $\Sigma(1385)$,
$\Sigma^0$, $\Xi$ as well as direct $\Lambda$ and heavier
resonances such as $\Omega$ are negligible. Since decay width of $\Sigma(1385)$ is
36-40 MeV, this contribution will give the $\Lambda$ source function an effectively
long lifetime and might influence  low $Q$ behavior of $C(Q)$, but will
not affect $\lambda$.  Thus, we do not treat daughters from
$\Sigma(1385)$ as a part of the long-lived parents but regards them as a
part of direct $\Lambda$. One needs to invoke
dynamical simulations for more serious estimates of these short-lived
resonance effects.

Then $N^p$ consists of contribution from $\Sigma^0$ and $\Xi$. 
The fraction of $\Sigma^0$ to $\Lambda$ in heavy ion collisions is not
experimentally known, because of the difficulty in reconstruction of
$\Sigma^0 \rightarrow \Lambda \gamma$ process. Here we utilize an
experimental result in $p+$Be collisions at $p_{\text{lab}}=28.5$GeV
\cite{Sigma0exp}, $N_{\Sigma^0}/N_{\Lambda}=0.278$. 
Although the production process of the hyperons in $pA$ collisions could
be different from that in heavy ion collisions, we note that this ratio
is consistent with thermal model calculations \cite{Redlich_private}.

$\Xi$ yields in Au+Au collisions at $\sqrt{s_{NN}}=200$
GeV has been shown to 15\% of total $\Lambda$ \cite{agakishiev12:_stran_enhan_in_cu_cu}. 
Here we assume that a part of $\Xi$ decay contribution to $\Lambda$ is
excluded by the candidate selection employed in the STAR measurement
\cite{STAR}, according to the distance of closest approach less than
0.4cm which is comparable to minimum decay length of $\Xi$. 
Combining these fractions of resonances to $\Lambda$ and 
subtracting the $\Xi$ contributions from $N_{\text{tot}}$,
We obtain $\lambda=(0.67)^2$. If we take account of the $\Xi$
contribution into the total yields, $\lambda=(0.572)^2$. Since
 the above selection does not exclude all of $\Xi$, the realistic value
 could be a little smaller than $(0.67)^2$, but larger than $(0.572)^2$.
We estimate the higher resonance contribution to $N_{\text{tot}}$ to within a
few percent. Therefore, we reanalyze the data by correcting the
correlation function obtained from the cylindrical source function
\eqref{eq:source}, with $\lambda=(0.67)^2$. 
We confirmed that the lower value $\lambda=(0.572)^2$ only gives
quantitative change in $\chi^2/N_{\text{dof}}$ values in the following
analyses.

\subsection{Effects on Correlation function}

\begin{figure*}
 \centering
 \includegraphics[width=\columnwidth]{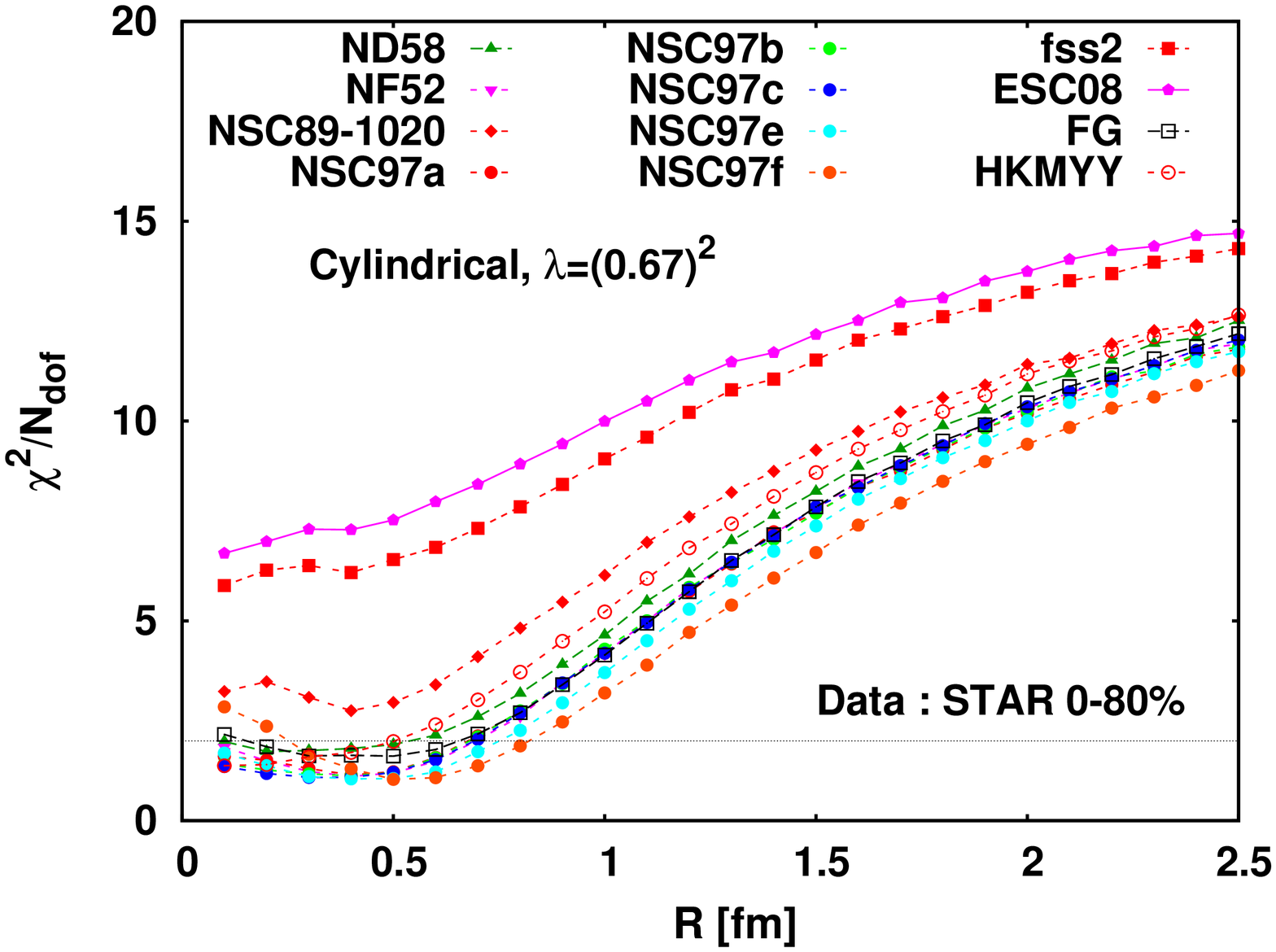}
 \includegraphics[width=\columnwidth]{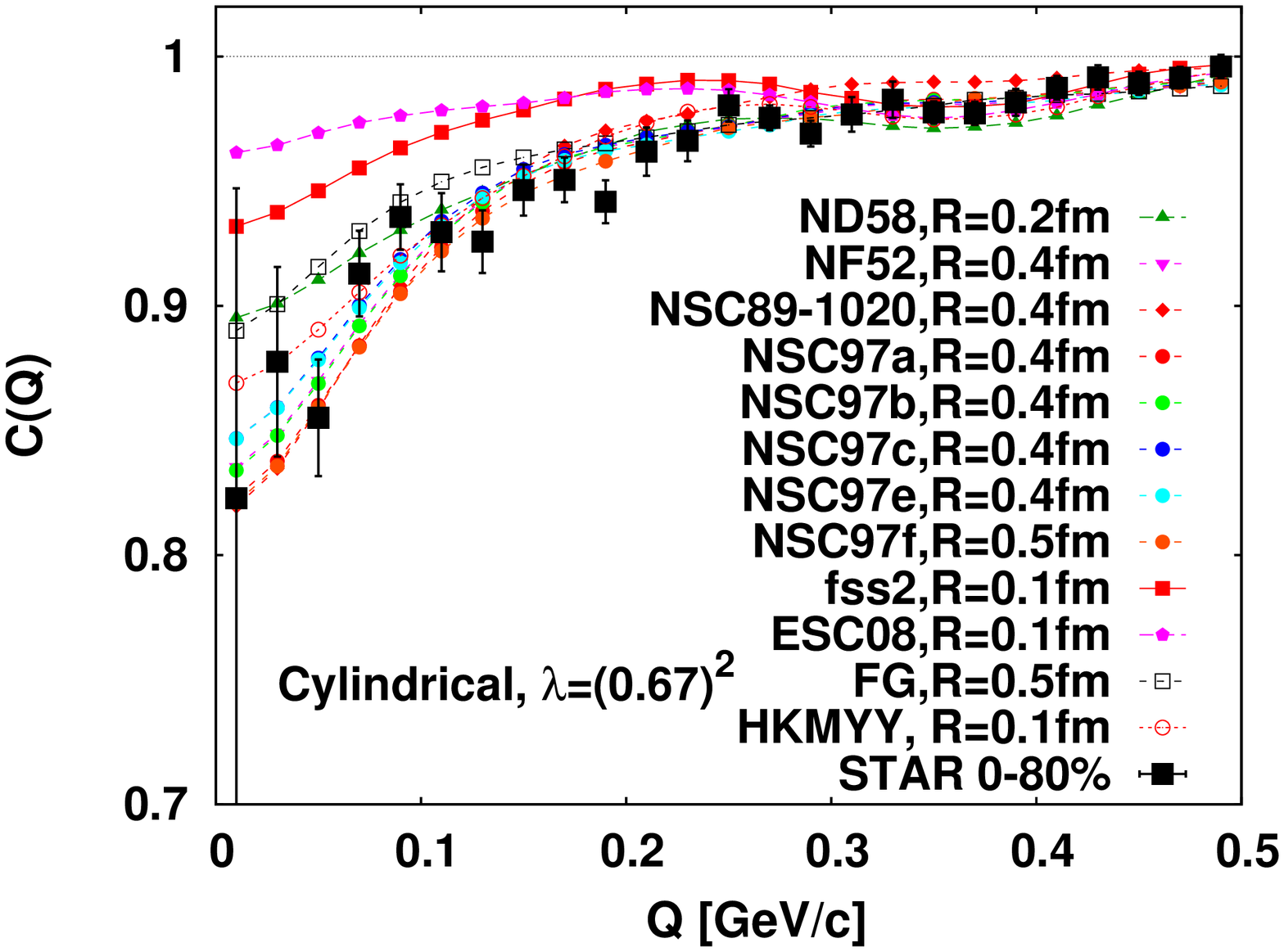}
 \caption{Same as Fig.~\ref{fig:cyli}, but corrected for $\Sigma^0$
 feed-down contribution by $\lambda=(0.67)^2$.}
 \label{fig:sigma_corrected}
\end{figure*}

Results of reanalysis with the feed-down contribution are shown in
Fig.~\ref{fig:sigma_corrected}. The correction factor $\lambda=(0.67)^2$
leads to $C(Q=0)=0.776$. This implies  the reduction of not only the
intercept, but also whole correlation function. As a result, $C(Q)$
becomes less sensitive to the difference among the potentials and
we have more potentials which can fit the data. For instance, NSC97
potentials now reproduce the data except for NSC97d. However, 
one immediately notes that  the potentials reproduce the data with
unphysically small transverse size, $R \leq 0.5$fm. Owing to the
multiplicative factor $\lambda$, the tail of $C(Q)$ becomes
closer to unity. Then smaller size is preferred to fit the long tail in
the STAR data. Thus. if we adopt the smaller value of $\lambda$, 
the minimum of $\chi^2$ shifts to smaller $R$ values. 
As explained in Sec.~\ref{sec:flow}, the smaller size is accompanied by
the stronger effect of the attraction. ND56 and NSC89-820,  which
already overshoot the data in the uncorrected case due to too strong
attraction in Fig.~\ref{fig:cyli}, can no longer reproduce the data
while some of those with weaker attaction fit the data well.  fss2 and
ESC08 overshoot the data despite the weaker attaction. Non-monotonic
behavior in $\chi^2/N_{\text{dof}}$ indicate that both may fit the low
$Q$ region with $R\simeq 0.4$fm, but minimum $\chi^2$ cannot be achived
due to the tail of $C(Q)$.

\subsection{Residual correlation}

The agreement between the data and models at small source size cannot be
physically reasonable result. It rather suggests that there exist
additional sources of the correlation which give the long tail of
$C(Q)$.  In Ref.~\cite{STAR}, a Gaussian term with two additional
fitting parameters $a_{\text{res}}$ and $r_\text{res}$,
\begin{equation}
 C_{\text{res}}(Q) = a_{\text{res}}e^{-r_{\text{res}}^2 Q^2},\label{eq:c_res}
\end{equation}
is employed in the fitting function $C_{\text{fit}}(Q)$ to account for
the long tail as a residual correlation, presumably caused by parent particles.
While this prescription is found to improve the quality of the fit,
the origin is not known. 
Here we investigate effects of the additional term by evaluating
$\chi^2$ as a function of $a_\text{res}$ and $r_{\text{res}}$ for each
of source size $R$.

 \begin{figure}
  \centering
  \includegraphics[width=\columnwidth]{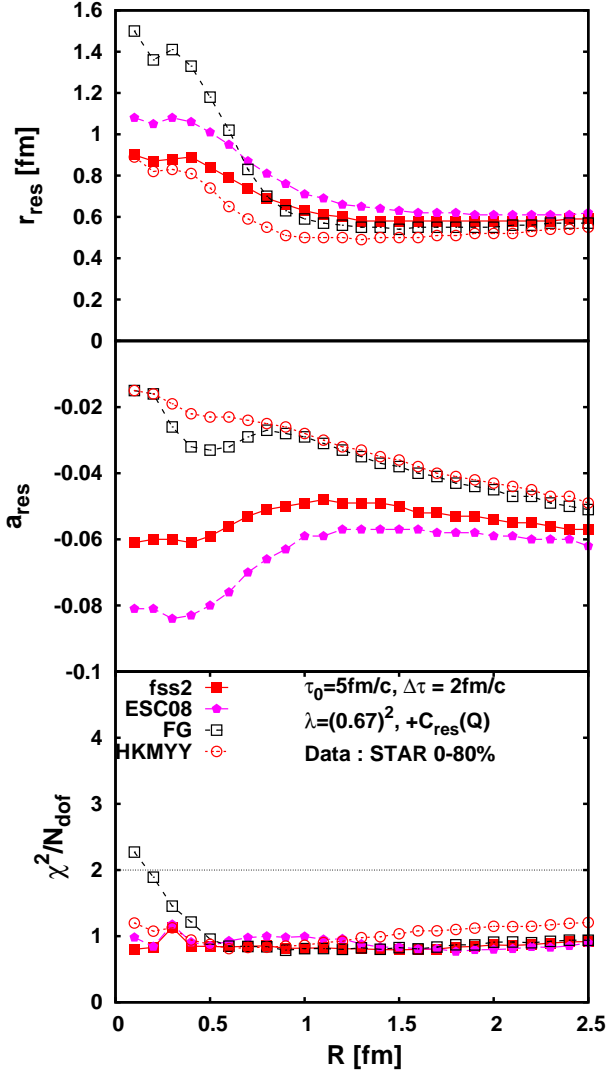}
  \caption{Residual correlation parameters as functions of $R$ for fss2,
  ESC08, FG and HKMYY potentials. Bottom:
  minimum of $\chi^2$. Top and middle panels : $r_{\text{res}}$ and
  $a_{\text{res}}$ giving the minimum $\chi^2$, respectively.}
  \label{fig:residual_etc}
 \end{figure}

Figure \ref{fig:residual_etc} shows results including the residual
correlation term for potentials which give
reasonable fits to the data before the feed-down correction.
One sees that $\chi^2/N_{\text{dof}}\simeq 1$ independent of the size
parameter for $R > 0.5$fm. The strength $a_{\text{res}}$ and the
spatial size parameter $r_{\text{res}}$ of the residual correlation exhibit $R$
dependence. In particular, for $R < 1$fm, $r_{\text{res}}$ becomes
larger as $R$ is decreased and eventually
$r_{\text{res}} > R$. On the other hand, $r_{\text{res}}$ and
$a_{\text{res}}$ for different potentials approach to common values as
$R$ increases. These tendencies might indicate a two-source structure in
the data.
Although unrealistic, in case of small $R$ and large $r_{\text{res}}$, 
the large variation in the strength $a_{\text{res}}$ (see the middle
panel of Fig.~\ref{fig:residual_etc}) indicates that the low $Q$
behavior is dominated by the residual term and the long tail of $C(Q)$
is fitted with the interaction and the collective effects in the source
function by small $R$. In this case, the role of each terms is inverted
but the general structure is kept such that the low $Q$ part is sensitive to the
interaction and the high $Q$ tail is attributed to a correlation in
a small size. The two source structure might
be more natural for small $r_{\text{res}}$ and large $R$, in which 
$r_{\text{res}}$ approaches to $0.6$fm and $R$ is comparable to
source sizes extracted from proton-proton correlation measurements. 

\begin{figure}
 \centering
 \includegraphics[width=\columnwidth]{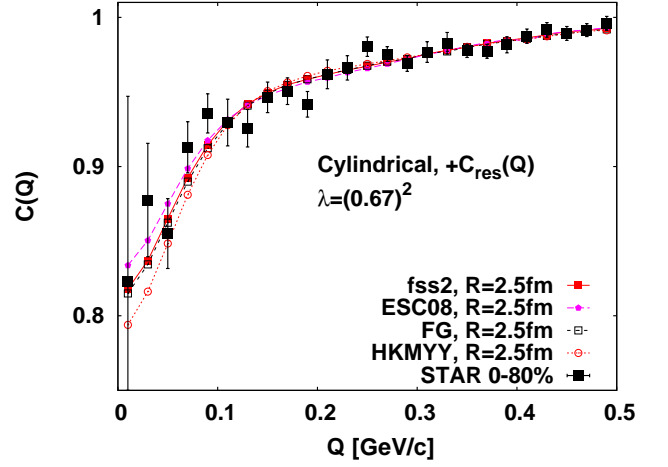}
 \caption{Correlation functions combined with the residual term and the
 feed-down correction for $R=2.5$fm. }
 \label{fig:c2_residual_sigma}
\end{figure}

We find that the same behavior is also seen in other potentials with
$1/a_0 \leq -0.8$fm. The feed-down correction reduces the sensitivity to
$\Lambda\Lambda$ interaction in low $Q$ region. Consequently, the data no
longer constraints the effective range,  as  demonstrated in
Fig.~\ref{fig:c2_residual_sigma}, where the correlation
functions are plotted for $R=2.5$ fm, in accordance with the $\chi^2$ analysis in Fig.~\ref{fig:residual_etc}.
All the shown potentials fit the data well and no
difference is seen in $Q > 0.15$GeV/$c$. At low $Q$, albeit small,
$C(Q)$ exhibit interaction dependence fairly reflecting the strength of
attraction, similarly to the right panel of Fig.~\ref{fig:cyli}. Note,
however, that FG and fss2 have a factor of two different effective range
while they have almost the same scattering length. We also find that all
the NSC97 potentials, of which $r_{\text{eff}}$ is broadly ranged from
1.15fm to 16.33fm and the scattering length is $1/a_0 < -2$fm$^{-1}$, can
reproduce the data with $\chi^2/N_{\text{dof}} \simeq 1$. 

In general, results with the residual correlation term \eqref{eq:c_res} depends on
the feed-down contribution $\lambda$. Since the value $\lambda=(0.67)^2$
takes only $\Sigma^0$ into account, this serves a minimal correction
owing to possible $\Xi$ contribution. We confirmed
the present result for the constraint on the scattering length, 
$1/a_0 < -0.8$fm,  holds for smaller $\lambda$ by repeating the same
analyses for $\lambda=(0.572)^2$ in which $\Xi$ contribution is
included. 

The above discussion applies to all the potentials with $a_0 < 0$
analyzed here. We note that there are two exceptions. We find that ND46
and NF42, which have the positive largest $1/a_0$ thus have a bound
state, can fit the data when  $R< 0.7$fm with the residual correlation taking
$2 < r_{\text{res}}  < 4$fm and $-0.2 < a_{\text{res}} <  -0.08$. 
We consider it to be coincidence, since it is accompanied with
$r_{\text{res}}$ larger than the source size and 
$\chi^2/N_{\text{dof}} \sim 1$ is achived only in small $R$ region.  As
we shall discuss below, the appearance of the bound state
should lead to suppression of $C(Q)$ at low $Q$ when the source size is
larger than $a_0$. Therefore, one may be able to confirm or rule out
this possibility by analyzing data of more central collisions, which are
expected to have a larger source size.

\section{Discussion}
\label{sec:discussion}

\subsection{Possible signal of $H$ resonance}

On the basis of the scattering length and the effective range of
the $\Lambda\Lambda$ interaction obtained in the present analyses, the
existence of $H$ particle as a bound state of \LL\ is not preferred.
This can be understood from the enhanced \LL\ correlation function
observed in the data compared with the free case.
If we had a bound state in \LL,
the correlation function would be suppressed from the free case.
The scattering wave function has the asymptotic form,
$\chi_q(r)=e^{-i\delta}\sin(qr+\delta)/qr$, where $q=Q/2$
is the relative momentum of $\Lambda$. 
In the case of small enough interaction range compared
with the source size, we can substitute the asymptotic form 
for the scattering wave function $\chi_q(r)$ in Eq.~\eqref{eq:c2_stat} 
and obtain the low energy limit of the
correlation function,
\begin{align}
C(Q) &\to \frac12 - \frac{1}{\sqrt{\pi}}\,\frac{a_0}{R} 
+ \frac14 \left(\frac{a_0}{R}\right)^2
\quad (Q \to 0)\ ,
\label{eq:lowE}
\end{align}
where the phase shift is given approximately as $\delta \simeq -a_0 q$.
For \LL\ interaction with a bound state ($a_0>0$),
The scattering wave function has a node at $r \simeq a_0$ at low energies,
then the correlation function is suppressed
compared with the free case in the low energy limit,
as long as the second term dominates in Eq.~\eqref{eq:lowE}.
Thus we would see suppressed $Q$ region if we have a bound state.
In practice, the interaction range is not small enough compared with the
source size considered here, thus the above estimate might not be precise.
It should be noted that the above argument is not valid,
when \LL\ is not the dominant component of $H$.

\begin{figure}[ht]
\includegraphics[width=3.375in]{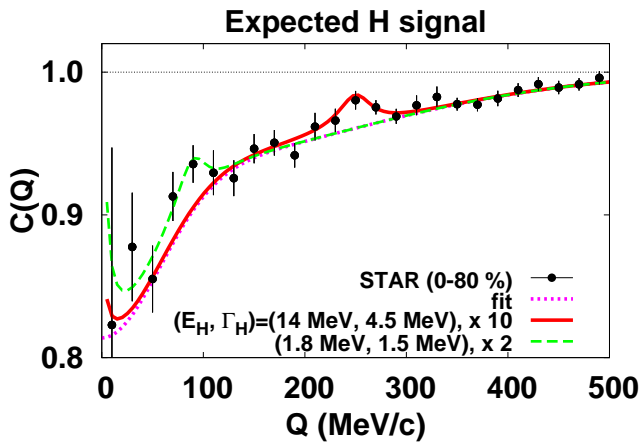}
\caption{Possible resonance $H$ signal in the \LL\ correlation function.
Signal for $(E_H, \Gamma_H)=(14~\MeV, 4.5~\MeV)$
and $(E_H, \Gamma_H)=(1.8~\MeV, 1.5~\MeV)$ are multiplied by 10 and 2,
respectively.}
\label{fig:H}
\end{figure}

The existence of $H$ as a resonance pole above the \LL\ threshold
is another interesting possibility,
as suggested in KEK experiments~\cite{E224,E522}.
While \LL\ potentials considered here do not have $H$
as an $s-$wave resonance, a quark model calculation
with instanton induced interaction allow the existence of resonance $H$
below the $\Xi N$ threshold~\cite{Takeuchi1991}.
In order to evaluate the strength of the resonance $H$ signal
in the correlation function, we have invoked the statistical model results.
In the statistical model~\cite{ExHIC}, $H$ ($\Lambda$) yield is calculated 
to be $N_H \simeq 1.3\times 10^{-2}$ ($N_{\Lambda} \simeq 30$)
per event per unit rapidity.
We here assume that the resonance $H$ is produced in a different mechanism
from the \LL\ potential scattering.
We also assume that the mass of $H$ is distributed 
according to the Breit-Wigner function, then the contribution of resonance $H$
in the \LL\ relative momentum spectrum is given as
\begin{align}
\frac{dN_H}{dydQ}= N_H f_\mathrm{BW}(E_Q) \frac{dE_Q}{dQ} \ ,
\end{align}
where $f_\mathrm{BW}(E)=\Gamma_H/[(E-E_H)^2+\Gamma_H^2/4]/2\pi$ is
the Breit-Wigner function.
In Fig.~\ref{fig:H}, we show the strength of the resonance $H$ signal.
We have fitted the STAR data in a simple smooth function,
and added the ratio of $dN_H/dydQ$ to the thermal \LL\ distribution,
$dN_{\Lambda\Lambda}/dydQ=4\pi q^2 N_{\Lambda\Lambda}\, \exp(-q^2/2\mu T)/(2\pi\mu T)^{3/2}/2$,
where $q=Q/2$, $N_{\Lambda\Lambda}=N_\Lambda^2$ and $\mu=M_\Lambda/2$.
The signal is multiplied by a factor.
The resonance parameters and the multiplication factor are chosen
so as to fit the bump structure in the data 
as if the bump is the resonance $H$ signal.

We find that the resonance $H$ signal relative to the thermal background
is small and higher statistics is necessary to confirm its existence,
especially when the resonance energy is large, $E_H>10~\MeV$.
In order to demonstrate this point, we show the results with 
$(E_H, \Gamma_H)=(14~\MeV, 4.5~\MeV)$
and
$(E_H, \Gamma_H)=(1.8~\MeV, 1.5~\MeV)$,
where the signal is multiplied by 10 and 2, respectively.
In the case where the resonance energy is large,
the signal is in the same order of the bump height 
after multiplied by 10.
The bump structure around 14 MeV  ($Q \sim 250~\MeV/c$)
in the correlation function data for the 0-80 \% centrality 
seems to come from statistical fluctuations
and is not considered to be the signal of the resonance $H$,
since the position and height depend on the centralilty~\cite{Shah-private}.
If the resonance $H$ exists at small energy above the \LL\ threshold,
it may be possible to detect the signal.
The bump in the data at around 1.8 MeV ($Q \sim 90~\MeV/c$)
is also considered to come from the statistical fluctuation,
and its strength is about twice of the statistical model estimate 
of the $H$ signal. If we can reduce the error to a half,
it would be possible to confirm or rule out the existence of resonance $H$
at low energies.

\subsection{Implication to $\Lambda\Lambda{N}$ three-body interaction}

Another interesting implication of the present analyses
is the difference between the vacuum and in-medium \LL\ interactions.
The \LL\ interactions in Refs.~\cite{FG,Hiyama} 
may be less attractive than those expected from the \LL\ correlation data;
as seen in Figs.~\ref{fig:LL_correlation_static} and \ref{fig:cyli},
\LL\ correlations from these two interactions tend to be smaller than the data.
The strengths of these \LL\ interaction are fitted to the \LL\ bond energy
in the Nagara event
$\DBLL=1.01 \pm 0.20^{+0.18}_{-0.11}~\MeV$~\cite{Nagara}.
%
The \LL\ bond energy is recently updated 
to be a smaller value, $\DBLL=0.67 \pm 0.16~\MeV$
($B_{\Lambda\Lambda}=6.91~\MeV$)~\cite{Nagara-Update}
following the $\Xi^-$ mass update by the Particle Data Group~\cite{PDG2008}.
%
The updated $\DBLL$ value could imply further weaker \LL\ interaction.
For example, the HKMYY interaction was updated to fit the new
$\Delta B_{\Lambda\Lambda}$~\cite{HKYM2010},
and the scattering parameters become
$(a_0, r_\mathrm{eff})=(-0.44~\fm, 10.1~\fm)$,which is outside of
the favored region based on the analyses without the feed-down correction.

While it is still premature to draw any conclusion,
less attractive \LL\ interaction in nuclei than the interaction in vacuum
would suggest the density dependence of the \LL\ interaction.
A part of \LL\ attraction comes from the coupling with the $\Xi N$ channel
$\Lambda\Lambda \leftrightarrow \Xi N$,
since the threshold energy difference is small, 
$M_\Xi + M_N - 2 M_\Lambda \sim 28~\MeV$.
The Pauli blocking in the $\Xi N$ channel leads to reducing
the attraction in the \LL\ channel.
All the nucleon $0s$ states are occuplied 
in $^{~~6}_{\Lambda\Lambda}\mathrm{He}$,
the nucleon in the intermediate state has to be in the $p$ state,
and the coupling effect is suppressed.
Myint, Shinmura and Akaishi found that
the coupled channel Pauli suppression effect results in the reduction
of the bond energy in $^{~~6}_{\Lambda\Lambda}\mathrm{He}$
by 0.09, 0.43 and 0.88 MeV 
for the ND, NSC97e and NF coupling strength, respectively~\cite{Myint2002}.
Their result with NSC97e interaction $\DBLL=0.64~\MeV$
roughly coincides with the updated value,
but  their estimate of the scattering parameters
$(a_0, \reff)=(-0.5~\fm, 8.41~\fm)$ is outside the favored region
by the STAR dat in which the feed down correction is not taken
into account. 
It would be an
interesting issue whether $\Lambda\Lambda$ interaction with strong
coupling with $\Xi N$ channel can consistently explain the Nagara event
and the RHIC data.

The coupled channel Pauli suppression 
is found to be one of the important origins
to generate repulsive contribution in nuclear matter
at high density~\cite{Kohno2013}.
Thus further investigations of \LL\ correlation and double hypernuclei
would open a way to access the density dependence of the \LL\ interaction,
or the $\Lambda\Lambda N$ three-body interaction,
which would be important to understand the origin of the additional repulsion
required to support massive neutron stars 
with strange hadrons~\cite{Demorest}.

\subsection{Comparison with the previous work}

Before closing the section, we briefly comment on the analysis
by the STAR Collaboration in \cite{STAR}. Using the Lednick\'{y} and
Lyuboshitz analytical model, they found that the same data favor, albeit
weak, a repulsive interaction $a_0 > 0$ in contrast to our analysis 
suggesting weakly attractive interaction; 
the phase shift at low energies, $\delta \simeq -a_0 k$,
increases when $a_0<0$ as suggested in the present work,
while it decreases when $a_0>0$ as in the analysis by the STAR collaboration.
\footnote{Note that sign convention of
$a_0$ is different from that in the present paper.}

The main reason for
this discrepancy would come from the difference in the treatment of the
intercept parameter $\lambda$. In the analysis in \cite{STAR}, it is a
fitting parameter to obtain minimum $\chi^2$ while we fix it from yields
of parent particles decaying to $\lambda$. Their result 
$\lambda=0.18\pm 0.05^{+0.12}_{-0.06}$ is much smaller than ours,
and gives  $C(0)$ closer to unity.
A small value of $\lambda$ leads to a small bare correlation
$C_\mathrm{bare}(Q)$ at small $Q$ as found from Eq.~\eqref{Eq:corrbare},
and the low energy value would be less than the free value,
$C_\mathrm{bare}(Q\to0) < 0.5$, implying  a repulsive interaction.
This indicates the importance
to understand the $\lambda$ value
in the analysis of $\Lambda\Lambda$ correlation aiming
at extracting the interactions. 

\section{Concluding remarks}
\label{sec:conclusion}

We have studied the $\Lambda\Lambda$ correlation function in
relativistic heavy-ion collisions. 
By using a simple static source model, 
we illustrate how the correlation function is sensitive to differences
in interaction potentials. From fits to the measured experimental data,
it turned out that the favored potentials
have small negative scattering length and effective range around 4~fm.
Then we examined effects of the collective expansion on the behavior of
the correlation function by making use of a thermal source model
with the boost-invariant expansion along the collision axis and
transverse expansion fitted to $p_T$ spectrum. 
We point out that the strong expansion modifies the behavior of the
correlation function at small $Q$, but it remains sensitive to the
potential. In particular, it turns out that the same potentials as the
static source case are favored.
We have obtained a set of the potentials which
give reasonable fit to the data with a small transverse
size, under an assumption that feed down correction to the
correlation function is negligible.
Such potentials are characterized by the scattering length 
$-1.8~\mathrm{fm}^{-1} < 1/a_0 < -0.8~\mathrm{fm}^{-1}$ 
and the effective range
$3.5~\mathrm{fm} < r_\mathrm{eff} < 7~\mathrm{fm}$,
as represented by a thick shaded area in Fig.~\ref{fig:AR-sel}. 

In the above analysis, we treat the size parameter $R$ as a fitting parameter to
the experimental data. The obtained source size is found to be somewhat
smaller than the HBT radii for protons. Discussion on the collision
dynamics based on the size, as done in pion HBT studies, is beyond our
scope in this paper.
A possible reason might be small scattering cross
sections of $\Lambda$ with other particles,
which give the correlation function less sensitive to the later stage of the
collision process. Although the agreement of the favored potentials
between the static source model  and the boost-invariant source
model analyses seems to suggest irrelevance of the detailed dynamics of collisions, 
our findings should be confirmed by studies with more realistic source
models. 

We have also studied effects of feed down correction from
$\Sigma^0 \to \Lambda \gamma$ decay in analogy with effects of
long-lived particles
in the pion HBT. The reduction of the intercept
parameter $\lambda$ is found to reduce the sensitivity of the
correlation function to $\Lambda\Lambda$ interaction in low $Q$ and lead to unphysically
smaller source size to fit high $Q$ tail. To remedy this, we have
considered an additional Gaussian term in the correlation function and
found that the data is well described by the two source structure. We
confirmed that the low $Q$ part remains sensitive to the interaction
with coarse resolution such that the correlation function is no longer
sensitive to the effective range. The resultant constraint on the
scattering length is found to be $1/a_0 < -0.8$fm$^{-1}$. In this case,
the source size is roughly consistent with proton HBT radii, but the
long tail in $Q$ is attributed to the additional term of which origin is
not known.

Since this term can be attributed to a source with small size, 
it might suggest a different source of $\Lambda$ production such as
a two-step hadron production mechanism,
in which inhomogeneous matter is first formed and decays into hadrons later.
%
Nevertheless, the agreement of the favored potentials in this work with
the most recent $\Lambda\Lambda$ potential models (fss2 and ESC08)
indicate relevance of our study and demonstrates the feasibility
of using the $\Lambda\Lambda$ correlation function to extract the
interaction.
Furthermore, 
channel coupling 
needs to be taken care of
for a better understanding of data.
Preliminary analysis~\cite{Hyp2012} shows that the coupling
with $\Xi N$ channel is not significant
as long as the coupling potential is not very strong.
A more consistent treatment 
and data with higher statistics are desired to pin down the \LL\ interaction.

\acknowledgements

The authors would like to thank N.~Shah and H. Z. Huang
for providing them the STAR data.
They also would like to acknowledge T.~Rijken, S.~Shinmura,
Y.~Yamamoto, E.~Hiyama, A.~Gal, Y.~Akaishi and B.~M{\"u}ller
for helpful discussions.
K.M. would like to thank K.~Redlich for discussion on $\Sigma^0$
decay and hyperon yields in thermal models. He also acknowledges other
members of Institute of Theoretical Physics in University of Wroclaw for
discussion.
Numerical computations were carried out on SR16000 at YITP in Kyoto
university. 
This work is supported in part by the Grants-in-Aid for Scientific Research
from JSPS
(Nos.
          (B) 23340067, 
          (B) 24340054, 
          (C) 24540271
),
by the Grants-in-Aid for Scientific Research on Innovative Areas from MEXT
(No. 2404: 24105001, 24105008), 
by the Yukawa International Program for Quark-hadron Sciences,
by HIC for FAIR
and
by the Polish Science Foundation (NCN), under Maestro grant 2013/10/A/ST2/00106.

\end{document}